\documentclass[12pt,agums]{aguplus}
\usepackage{times}
\usepackage{mathptmx}
\usepackage{epsfig}
\usepackage[displaymath]{lineno}
\linenumbers*[1]
\usepackage{setspace}
\figmarkfalse
\sectionnumbers
\lefthead{KORENAGA} 
\righthead{SCALING OF PLATE TECTONICS}
\slugcomment{Submitted to {\it Journal of Geophysical Research}, April 2010; Revised, July 2010; accpeted, August 2010.}

\authoraddr{Jun Korenaga,
Department of Geology and Geophysics, P.O. Box, 208109, Yale
University, New Haven, CT 06520-8109. (jun.korenaga@yale.edu)}

\begin{document}           % End of preamble and beginning of text.

\title{Scaling of plate-tectonic convection with pseudoplastic
rheology}

\author{Jun Korenaga}
\affil{Department of Geology and Geophysics, 
Yale University, New Haven, CT}

\begin{abstract}
  The scaling of plate-tectonic convection is investigated by
  simulating thermal convection with pseudoplastic rheology and
  strongly temperature-dependent viscosity. The effect of mantle
  melting is also explored with additional depth-dependent
  viscosity. Heat-flow scaling can be constructed with only two
  parameters, the internal Rayleigh number and the lithospheric
  viscosity contrast, the latter of which is determined entirely by
  rheological properties. The critical viscosity contrast for the
  transition between plate-tectonic and stagnant-lid convection is
  found to be proportional to the square root of the internal Rayleigh
  number. The relation between mantle temperature and surface heat
  flux on Earth is discussed on the basis of these scaling laws, and
  the inverse relationship between them, as previously suggested from
  the consideration of global energy balance, is confirmed by this
  fully dynamic approach. In the presence of surface water to reduce
  the effective friction coefficient, the operation of plate tectonics
  is suggested to be plausible throughout the Earth history.
\end{abstract}

\begin{article}
\section{Introduction}

Simulating mantle convection with plate tectonics in a fully dynamic
manner has become popular in the last decade or so
\cite[e.g.,][]{bercovici03a}, and quite a few studies have been
published addressing a variety of problems, including the significance
of 3-D spherical geometry \cite[]{richards01,vanheck08,foley09}, the
role of history-dependent rheology
\cite[e.g.,][]{tackley00c,ogawa03,landuyt08}, the initiation of
subduction \cite[]{solomatov04,gurnis04}, and applications to other
terrestrial planets \cite[e.g.,][]{lenardic04,oneill07c,landuyt09b}.
Around the same time, interests in the initiation and evolution of
plate tectonics over the Earth history have grown considerably
\cite[e.g.,][]{mojzsis01,bleeker03,harrison05,stern05,korenaga06,vankranendonk07,oneill07b,condie08b,bradley08,harrison09,herzberg10}. Many
of previous numerical studies on plate-tectonic convection are,
however, exploratory in nature, and scaling laws relevant to such
geological questions are yet to be established. Given the lack of
consensus on why plate tectonics can take place on Earth to begin with
\cite[e.g.,][]{moresi98,gurnis00,bercovici03b,korenaga07c}, it may be
premature to discuss the scaling of plate-tectonic convection, but it
is nonetheless important to seek a strategy to bridge geology and
geodynamics by taking into account peculiar complications associated
with plate tectonics.

In this study, I attempt to derive the scaling of plate-tectonic
convection using the so-called pseudoplastic rheology
\cite[]{moresi98}, in which the strength of plates is controlled by
temperature-dependent viscosity as well as brittle failure. It is
known that, for this approach to be successful, the friction
coefficient for brittle deformation has to be at least one order of
magnitude lower than suggested by laboratory experiments. The presence
of pore fluid deep in the oceanic lithosphere is required to explain
such low friction, and because oceanic lithosphere is likely to be
very dry upon its formation by melting under mid-ocean ridges
\cite[]{hirth96,evans05}, it may appear to be difficult to justify the
pseudoplastic approach. The upper half of oceanic lithosphere,
however, can be pervasively fractured by thermal cracking, and in the
presence of surface water, the deep hydration of oceanic lithosphere
is possible \cite[]{korenaga07c}. In this mechanism, the strong
temperature dependency of mantle rheology actually enhances thermal
cracking. Another concern with the pseudoplastic rheology is that it
is determined only by the instantaneous stress state and does not have
any memory to simulate preexisting weakness, though this limitation is
not as grave as it may appear. With the thermal cracking hypothesis,
the stiffest part of oceanic lithosphere is continually damaged as it
ages, so preexisting weakness is globally distributed.  Also, whatever
the actual weakening mechanism would be, oceanic lithosphere is
eventually subducted (on the time scale of 100~Myr), and its memory of
weakness would keep being lost in the deep mantle. For the evolution
of oceanic lithosphere, therefore, the difference between
instantaneous rheology and history-dependent rheology is not expected
to be vital \cite[]{tackley00c}. In plate tectonics, convective heat
loss is dominated by that from oceanic plates, so even with simple
pseudoplastic rheology, we may still hope to capture the gross
characteristics of mantle convection relevant to the long-term
evolution of Earth.

The purpose of this paper is two-fold. First, I will investigate the
scaling of plate-tectonic convection with `standard' pseudoplastic
rheology, which is controlled by friction-based yield stress and
temperature-dependent viscosity. Though there exist a number of
numerical studies using this rheology, the temperature dependency of
mantle viscosity is fairly weak in most of these studies
\cite[e.g.,][]{moresi98,lenardic04,stein04,oneill07c}; the maximum
viscosity variation due to temperature dependency is usually $10^6$.
This may not seem to be low because it is high enough to put
convection in the stagnant-lid regime without pseudoplastic rheology.
In basally-heated convection, which is commonly adopted by those
previous studies, there is an important difference regarding thermal
structure between stagnant-lid convection and plate-tectonic
convection. In the stagnant-lid regime, most of temperature variations
are taken up by the top thermal boundary layer, so viscosity variation
across the top boundary layer is close to the maximum viscosity
variation employed. In the plate-tectonic regime with basal heating,
the top and bottom thermal boundary layers have similar temperature
contrasts, i.e., the temperature contrast across the top boundary
layer (or plates) is basically halved, with the corresponding
viscosity variation of only $10^3$. As explained in more detail later
(\S\ref{sec:rheology}), the viscosity contrast across oceanic
lithosphere due to temperature dependency is expected to be at least
$\exp(20) \sim 5 \times 10^8$. It is important to use strongly
temperature-dependent viscosity so that we can discuss the scaling of
plate-tectonic convection with more confidence. The second objective
of this paper is to discuss the effects of mantle melting by adding
depth-dependent viscosity to the standard pseudoplastic rheology. How
mantle melting could modify the scaling of plate tectonics has
important implications for the thermal evolution of Earth
\cite[]{korenaga03a,korenaga06}, but this issue has not been
quantified by fully dynamic calculations.

This paper is organized as follows. After describing the details of
theoretical formulation (\S\ref{sec:formulation}), I will present
numerical results, together with scaling analysis to understand the
systematics of model behavior (\S\ref{sec:results}).  In the
discussion section (\S\ref{sec:discussion}), I will briefly explore
how new scaling laws may be used to inter when plate tectonics
initiated on Earth and how it evolved subsequently. Critiques on
previous attempts to derive the scaling of plate tectonics are also
provided.

\section{Theoretical Formulation \label{sec:formulation}}

\subsection{Mantle Rheology \label{sec:rheology}}

For temperature-dependent viscosity, I employ the following
linear-exponential form:
\begin{linenomath*}\begin{equation}
  \eta^*_T = \exp [\theta (1-T^*)],
\label{eq:viscT}
\end{equation}\end{linenomath*}
where viscosity is normalized by reference viscosity $\eta_0$
defined at $T^*=1$. Temperature is normalized as
\begin{linenomath*}\begin{equation}
T^* = \frac{T-T_s}{\Delta T},
\end{equation}\end{linenomath*}
where $T_s$ is the surface temperature ($\sim$273~K), and $\Delta T$
is the (arbitrary) temperature scale. The degree of temperature
dependency is controlled by the Frank-Kamenetskii parameter $\theta$,
which can be related to the activation energy $E$ as
\cite[e.g.,][]{solomatov00a}
\begin{linenomath*}\begin{equation}
  \theta = \frac{E \Delta T}{R(T_s+\Delta T)^2},
\label{eq:theta}
\end{equation}\end{linenomath*}
where $R$ is the universal gas constant.  For $E$ of
$\sim$300~kJ~mol$^{-1}$ \cite[e.g.,][]{karato93} and $\Delta T$ of
$\sim$1300~K, for example, $\theta$ is $\sim$20.

The coldest part of the lithosphere would be very stiff due to this
strongly temperature-dependent viscosity, but it can also deform by
brittle failure. In the continuum limit, this brittle behavior can be
modeled by nonlinear effective viscosity that is adjusted to ensure
the stresses remain bounded by the yield stress envelope
\cite[]{moresi98}. The yield stress criterion for brittle deformation
may be expressed as
\begin{linenomath*}\begin{equation}
\tau_y = c_0 + \mu \rho_0 g z,
\end{equation}\end{linenomath*}
where $c_0$ is the cohesive strength, $\mu$ is the friction
coefficient, $\rho_0$ is reference density, $g$ is gravitational
acceleration, and $z$ is depth.  Using the length scale $D$, which is
the depth of a fluid layer, and the stress scale $\eta_0 \kappa /
D^2$, where $\kappa$ is thermal diffusivity, the criterion may be
nondimensionalized as
\begin{linenomath*}\begin{equation}
\tau^*_y = \tau^*_0 + \tau^*_1 z^*,
\end{equation}\end{linenomath*}
where 
\begin{linenomath*}\begin{equation}
\tau^*_0 = \frac{c_0 D^2}{\kappa \eta_0},
\end{equation}\end{linenomath*}
and
\begin{linenomath*}\begin{equation}
\tau^*_1 = \frac{\mu \rho_0 g D^3}{\kappa \eta_0}.
\end{equation}\end{linenomath*}
Using the Rayleigh number defined as
\begin{linenomath*}\begin{equation}
Ra = \frac{\alpha \rho_0 g \Delta T D^3}{\kappa \eta_0},
\label{eq:Ra}
\end{equation}\end{linenomath*}
where $\alpha$ is thermal expansivity, the criterion
can also be expressed as
\begin{linenomath*}\begin{equation}
\tau^*_y = \tau^*_0 + \gamma Ra \, z^*,
\end{equation}\end{linenomath*}
where 
\begin{linenomath*}\begin{equation}
\gamma = \frac{\mu}{\alpha \Delta T}.
\label{eq:gammadef}
\end{equation}\end{linenomath*}
In this study, the cohesive strength is assumed to be negligibly small
compared to the depth-dependent component, and $\tau^*_0$ is set to
$\tau^*_1 \times 10^{-5}$. This is a reasonable approximation given
experimental data on rock friction at low hydrostatic pressure
\cite[e.g.,][]{byerlee78} and also allows me to focus on the single
parameter $\gamma$.  Note that a nonzero cohesive strength term
appearing for experimental data at high confining pressures could
arise from a pressure-dependent friction coefficient with zero
cohesive strength.

The nonlinear effective viscosity for the plastic deformation is 
calculated as
\begin{linenomath*}\begin{equation}
\eta^*_y = \frac{\tau^*_y}{e^*_{\mbox{\scriptsize II}}},
\end{equation}\end{linenomath*}
where $e^*_{\mbox{\scriptsize II}}$ is the second invariant of the
(nondimensionalized) strain rate tensor. The transition between
plastic and ductile deformation is handled by using the harmonic mean
of the temperature-dependent viscosity and the above effective
viscosity as
\begin{linenomath*}\begin{equation}
\eta^* = \left(\frac{1}{\eta^*_T} + \frac{1}{\eta^*_y}\right)^{-1}. 
\label{eq:visc0}
\end{equation}\end{linenomath*}
The effective viscosity for plastic deformation $\eta^*_y$ is
calculated for any deformation, but when stresses are smaller than the
yield stress (i.e., $\eta^*_y$ is large), the harmonic mean above
will be dominated by $\eta^*_T$. 

The linear-exponential form of temperature-dependent viscosity
[equation~(\ref{eq:viscT})] predicts much smaller viscosity variation
across the entire lithosphere than the more realistic Arrhenius form,
$\exp(E/RT)$, but because the above pseudoplastic rheology effectively
eliminates a drastic viscosity increase in the upper half of the
lithosphere, the difference between the linear-exponential and
Arrhenius forms is actually small \cite[]{solomatov04}. It is still
important to use the realistic value of $\theta$ as it controls the
strength of the lower half of the lithosphere.

When considering the effects of mantle melting, I will add
depth-dependent viscosity as
\begin{linenomath*}\begin{equation}
\eta^* = \left(\frac{1}{\eta^*_T \, Z(z^*)} + \frac{1}{\eta^*_y}\right)^{-1},
\label{eq:visc1}
\end{equation}\end{linenomath*}
where
\begin{linenomath*}\begin{equation}
Z(z^*) = \left\{ 
\begin{array}{cl} \Delta \eta & \mbox{for} \,\,\,\, z^* \le h^* \\
  1 & \mbox{for} \,\,\,\, z^* > h^*  \end{array} \right.
\label{eq:zdep}
\end{equation}\end{linenomath*}
where $h^*$ is the thickness of dehydrated mantle and $\Delta \eta$ is 
a viscosity contrast introduced by dehydration.

\subsection{Governing Equations and Heating Mode\label{sec:heatingmode}}

The nondimensionalized governing equations for thermal convection of
an incompressible fluid consist of the conservation of mass,
\begin{linenomath*}\begin{equation}
\nabla \cdot {\bf u^*} = 0,
\label{eq:mass}
\end{equation}\end{linenomath*}
the conservation of momentum,
\begin{linenomath*}\begin{equation}
-\nabla P^* 
+ \nabla \cdot [\eta^* (\nabla {\bf u^*} + \nabla{\bf u^*}^T)]
- Ra T^* {\bf e}_z = 0,
\label{eq:momentum}
\end{equation}\end{linenomath*}
and the conservation of energy,
\begin{linenomath*}\begin{equation}
\frac{\partial T^*}{\partial t^*}
+ {\bf u^*} \cdot \nabla T^* = 
\nabla^2 T^* + H^*.
\label{eq:energy}
\end{equation}\end{linenomath*}
The unit vector pointing downward is denoted by ${\bf e}_z$.  The
spatial coordinates are normalized by the length scale $D$, and time
is normalized by the diffusion time scale, $D^2/\kappa$,. Velocity
${\bf u^*}$ is thus normalized by $\kappa/D$. Dynamic pressure $P^*$
and heat generation $H^*$ are normalized by $\eta_0 \kappa / D^2$ and
$k \Delta T / (\rho_0 D^2)$, respectively, where $k$ is thermal
conductivity.

In this study, I will focus on thermal convection that is purely
internally heated, by using the insulated bottom boundary condition.
There will be no thermal boundary layer at the bottom, simplifying the
scaling analysis of numerical results. This heating mode is also
appropriate for the majority of the Earth history
\cite[][section~5.1]{korenaga08d}.  At the same time, a temperature contrast across
the fluid layer is not known {\it a priori}, so the maximum
temperature, $T^*_{\mbox{\scriptsize max}}$, is not guaranteed to be
unity. Some {\it a posteriori} rescaling is thus necessary. The
Frank-Kamenetskii parameter is recalculated from its original value
$\theta_0$ as
\begin{linenomath*}\begin{equation}
\theta = \theta_0 \, T^*_{\mbox{\scriptsize max}},
\end{equation}\end{linenomath*}
so that $\exp(\theta)$ corresponds to the actual maximum viscosity
variation due to temperature dependency. The internal Rayleigh number
may also be defined with $T^*_{\mbox{\scriptsize max}}$ as
\begin{linenomath*}\begin{equation}
Ra_i = Ra \, T^*_{\mbox{\scriptsize max}} 
\exp[\theta_0 ( T^*_{\mbox{\scriptsize max}}-1)],
\end{equation}\end{linenomath*}
in which the total temperature contrast is $T^*_{\mbox{\scriptsize
    max}} \Delta T$, and the internal viscosity, $\eta_i$, is assumed
to be $\eta_0 \exp[\theta_0 (1- T^*_{\mbox{\scriptsize max}})]$.
Because of purely internal heating, the surface heat flux $q$ is, at a
statistical equilibrium, equal to total heat generation in the fluid
divided by surface area
\begin{linenomath*}\begin{equation}
q = \rho_0 D H,
\end{equation}\end{linenomath*}
and the corresponding Nusselt number is calculated as
\begin{linenomath*}\begin{equation}
Nu = \frac{q}{k T^*_{\mbox{\scriptsize max}} \Delta T/D}
= \frac{H^*}{T^*_{\mbox{\scriptsize max}}}.
\end{equation}\end{linenomath*}

The internal heating ratio (IHR), $\xi$, is the difference between
heat flux out of the top boundary and that into the bottom boundary,
normalized by the former \cite[e.g.,][]{mckenzie74}, i.e.,
\begin{linenomath*}\begin{equation}
    \xi = \frac{Nu_{\mbox{\scriptsize top}} - Nu_{\mbox{\scriptsize bot}}}{Nu_{\mbox{\scriptsize top}}},
\end{equation}\end{linenomath*}
and because the bottom boundary is insulated in this study
($Nu_{\mbox{\scriptsize bot}}$=0), IHR is unity for all runs as long
as $H^*$ is positive. The internal heat production $H^*$ does not
directly correspond to the amount of radiogenic heat production in the
mantle, which may be referred here as $H^*_{\mbox{\scriptsize
    rad}}$. Over the Earth history, the mantle has been (usually)
cooling with time \cite[]{abbott94,herzberg10}, and in the study of
mantle convection, this secular cooling is often included as part of
`internal' heating. So $H^*$ represents both radiogenic heat
production and secular cooling. Secular cooling is a transient
phenomenon, and directly simulating it requires us to assume an
initial condition for subsolidus mantle convection on Earth, which is
hardly known.  Numerical models for mantle convection are therefore
typically run for a number of convective overturns to reach a
statistical equilibrium so that model results do not strongly depend
on employed initial conditions. This steady-state modeling approach
has to include secular cooling as part of internal heating, in order
to simulate an Earth-like IHR. The thermal evolution of Earth can be
studied reasonably well by assuming that the mantle is in a quasi
steady state at each time step \cite[e.g.,][]{daly80}.

It is important to distinguish IHR from the convective Urey ratio,
$Ur$, which is the ratio of radiogenic heat production in the mantle
over the mantle heat flux \cite[]{christensen85}, i.e.,
\begin{linenomath*}\begin{equation}
    Ur = \frac{H^*_{\mbox{\scriptsize rad}}}{Nu_{\mbox{\scriptsize top}}}.
\end{equation}\end{linenomath*}
The Urey ratio is directly related to the chemical composition of
Earth's mantle, and it is a key parameter to describe the thermal
budget of Earth. When $Ur$ is discussed, radiogenic heat production and
secular cooling are considered separately. As noted by
\cite{korenaga08d}, there has been some misunderstanding in the
literature by confusing $Ur$ with IHR or by underestimating the
significance of secular cooling, and unfortuately, such confusion
still seems to continue \cite[e.g.,][]{deschamps10}. IHR can be
related to the convective Urey ratio as \cite[]{korenaga08d}
\begin{linenomath*}\begin{equation}
    \xi \approx 1 - \frac{C_c}{C_m+C_c} (1-Ur),
\end{equation}\end{linenomath*}
where $C_m$ and $C_c$ are, respectively, the heat capacities of the
mantle and the core. The present-day Urey ratio is probably $\sim$0.2
\cite[]{korenaga08d}, but because the core heat capacity is only
$\sim$1/5 of the whole Earth value, the present-day IHR for Earth's
mantle is estimated to be $\sim$0.9 \cite[]{korenaga08d}. Based on
thermal history considerations, the Urey ratio may have been higher in
the past \cite[]{korenaga06,herzberg10}, so IHR is likely to have been
closer to unity than at present. To first order, therefore, the use of
purely internal heating ($\xi$=1) appears to be a reasonable
simplification.

\subsection{Notes on Modeling Strategy \label{sec:strategy}}

Besides the use of pseudoplastic rheology, the numerical model of
mantle convection as specified in the previous sections is kept simple
to facilitate the interpretation of modeling results, and the
potential significance of realistic complications, which are neglected
in this study, are discussed in the following.

Because of the insulating boundary condition, bottom heat flux is
zero, so there are no upwelling plumes in the model. The influence of
plumes on plate dynamics thus cannot be examined. Because of the
nearly unity IHR expected for Earth's mantle
(\S\ref{sec:heatingmode}), however, such influence may not be of first
order. The governing equations employed are based on the Boussinesq
approximation \cite[e.g.,][]{schubert01}, so adiabatic gradients are
zero (i.e., the total temperature contrast $T^*_{\mbox{\scriptsize
    max}} \Delta T$ is the superadiabatic temperature contrast), and
the model temperature corresponds to potential temperature. The
effects of compressibility on the gross characteristics of thermal
convection have been known to be rather minor
\cite[]{jarvis80,bercovici92}.

For the ductile deformation of the mantle, the Newtonian rheology with
linear-exponential temperature dependency is adopted
[equation~(\ref{eq:viscT})], but mantle rheology is known to be much
more complex depending on, at least, stress, pressure, grain size, and
chemical composition \cite[e.g.,][]{karato93}. In case of
pseudoplastic rheology, the difference between the Arrhenius rheology
and its linear-exponential approximation is not important as already
mentioned, and I choose to use the latter because it is specified by
only one nondimensional parameter $\theta$, whereas the Arrhenius-type
temperature dependency requires three \cite[e.g.,][]{korenaga09a}.
Non-Newtonian, stress-dependent rheology can be approximated by
Newtonian rheology if the activation energy is properly scaled
\cite[]{christensen84c}. The importance of pressure dependence caused
by the activation volume is not clear at the moment. First of all,
activation volumes for mantle rheology are still poorly known even for
upper mantle minerals \cite[]{korenaga08a}. Second, viscosity increase
with increasing pressure should be at least partly cancelled by
viscosity decrease with increasing temperature along the mantle
adiabat. With the Boussinesq approximation employed here, the use of
pressure-independent rheology actually requires non-zero activation
volume, the effect of which is assumed to be cancelled exactly by
temperature variations along the adiabat. Grain size variation can
affect mantle dynamics considerably \cite[e.g.,][]{solomatov96b}, but
how grain size should evolve in the convecting mantle is still poorly
understood, so it appears premature to consider its effect in this
study. The effect of composition on mantle rheology is taken into
account when dehydration stiffening is effected by depth-dependent
viscosity [equation~(\ref{eq:zdep})].  There are of course other
compositional effects \cite[e.g.,][]{karato08}, but the effect of
dehydration appears to be most important at least for the upper mantle
rheology \cite[e.g.,][]{karato86a,mei00a,mei00b,faul07}, and mantle
dehydration is always expected whenever mantle melts \cite[]{hirth96}.

Another important rheological aspect for large-scale mantle dynamics
is a viscosity jump at the base of the upper mantle, which has been
estimated to be on the order of $\sim$10-100 primarily through the
geodynamical modeling of Earth's geoid \cite[e.g.,][]{hager84}. Such
inference is, however, also known to suffer from considerable
nonuniqueness \cite[e.g.,][]{king95,kido97}, and the viscosities of
the upper and lower mantle may not be very different if the mantle
transition zone has a lower viscosity \cite[]{soldati09}. Furthermore,
even if the lower mantle does have a higher viscosity than the upper
mantle, it applies only for the present-day situation. When the mantle
was hotter in the past, the viscosity contrast may be smaller or even
reversed if the lower mantle rheology is more temperature-dependent
(i.e., higher activation energy) than the upper mantle counterpart.
Rheological stratification in the mantle is an important subject, but
these uncertainties imply a variety of situations to be considered, so
it is left for future studies.

The mantle transition zone is also characterized by multiple phase
transitions, and in particular, the effects of the endothermic phase
change at the base of the transition zone on large-scale mantle
circulation was once a popular topic in geodynamics
\cite[e.g.,][]{christensen84a,tackley93,solheim94a,yuen94}. Numerical
studies with strong plates exhibit, however, only a modest influence
of endothermic phase change on mantle dynamics
\cite[e.g.,][]{zhong94}, and recent experimental studies further
suggest that the Clapeyron slope of the endothermic phase change is
likely to be only $-1.3$~MPa~K$^{-1}$ \cite[]{katsura03,fei04}, which
is much less negative than previously thought. Modeling phase
transitions, therefore, is not considered to be essential. 

Finally, the model is 2-D Cartesian, whereas the use of a 3-D
spherical shell would be most appropriate. The restriction to 2-D
modeling is primarily to generate a large number of modeling results
(with modest computational resources) so that scaling analysis becomes
more robust, though I do not expect scaling laws to change drastically
by moving from 2-D to 3-D. The effect of sphericity would likely be
of minor nature \cite[]{bercovici00}. Based on isoviscous convection
models using 3-D spherical shells, for example, \cite{deschamps10}
recently derived the following heat-flow scaling (adapted here for the
case of purely internal heating),
\begin{linenomath*}\begin{equation}
    Nu \approx 0.59 f^{0.05} Ra^{0.300-0.003f},
\end{equation}\end{linenomath*}
where $f$ is the ratio of the core radius to the total radius of a
planet. The ratio $f$ is 0.55 for Earth and unity for Cartesian, so it
can be seen that sphericity has virtually no impact on this scaling. 

The convection model of this study is, therefore, simple but probably
not simpler than necessary. In any event, this study should provide a
reference point, by which the effects of any additional complication
can be quantified in future.

\section{Numerical Results and Scaling Analysis \label{sec:results}}

The finite element code of \cite{korenaga03b} was used to solve the
coupled Stokes flow and thermal advection-diffusion
equations~(\ref{eq:mass})-(\ref{eq:energy}). The benchmark tests of
this code can be found in \cite{korenaga03b} for Newtonian rheology
and in \cite{korenaga09a} for non-Newtonian rheology. To reduce wall
effects, the aspect ratio of the convection model is set to 8, and the
model domain is discretized with $400 \times 50$ uniform 2-D
quadrilateral elements. With this mesh resolution, model parameters
are chosen so that $Nu$ does not exceed 20 and the top thermal
boundary layer contains at least a few elements vertically on average.
The nondimensional surface temperature is fixed to zero, and the
bottom boundary is insulated.  The top and bottom boundaries are
free-slip, and a reflecting boundary condition is applied to the side
boundaries. In all cases, $Ra$ is set to $10^6$, but $Ra_i$ varies
greatly because of different combinations of $\theta_0$ and $H^*$ (and
thus $T^*_{\mbox{\scriptsize max}}$).

The initial temperature condition is specified as
\begin{linenomath*}\begin{equation}
T^*(x^*,z^*) = z^*+ a \cos(\pi x^*) \sin (\pi z^*) + \epsilon,
\end{equation}\end{linenomath*}
where $a$ is usually 0.2, and $\epsilon$ is random fluctuation with
the amplitude of $10^{-3}$. When the assumed mantle rheology is
appropriate for the operation of plate-tectonic convection, this
initial condition quickly brings the system to that mode of
convection. Otherwise, the system gradually migrates into the mode of
stagnant-lid convection. If I start with a uniformly hot fluid
instead, the system always begins with stagnant-lid convection, and
sublithospheric mantle is heated up considerably until the onset of
plate tectonics. Very low viscosity (and thus very high convective
velocity) beneath the stagnant lid during this initial period means
exceedingly small time steps for numerical integration, so this type
of initial condition is not computationally efficient when aiming at
statistically steady states required for scaling analysis.

\subsection{Convection Diagnostics}

\begin{figure*}[t]
\small
\singlespacing
\begin{center}
\vspace*{-2cm}
\epsfig{file=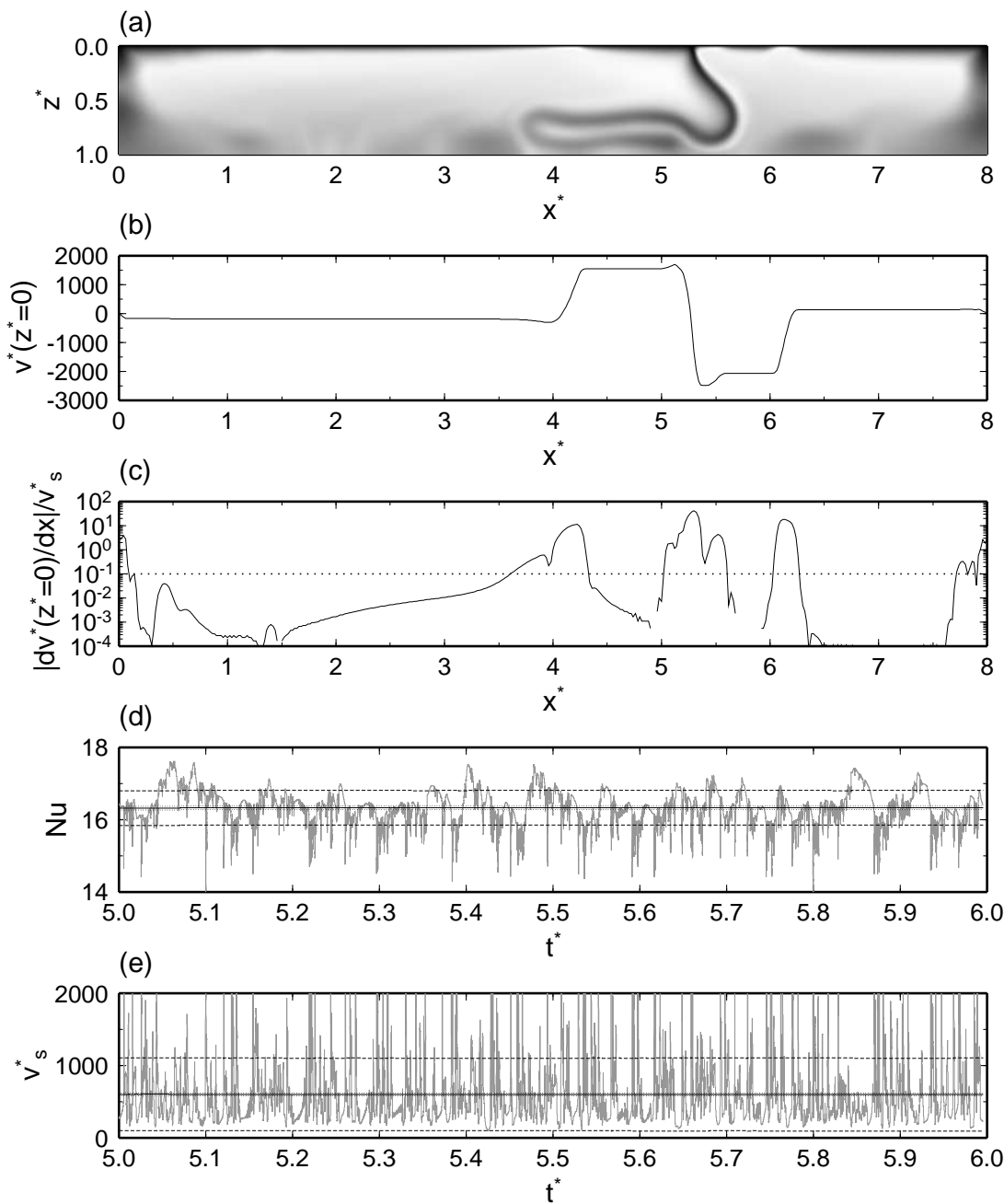,height=18cm}
\end{center}
\vspace{-1cm}
\caption{Example of simulation results from the case of
  $\gamma$=0.6, $\theta_0$=15, and $H^*$=20 with the standard
  pseudoplastic rheology [equation~(\ref{eq:visc0})].  (a) Snapshot of
  the temperature field. Darker shading corresponds to lower
  temperature. (b) Surface velocity profile and (c) horizontal strain
  rate scaled by by the average surface velocity, corresponding to the
  snapshot shown in (a). $P_{0.1}$ is 0.76 for this particular
  velocity profile, i.e., 76\% of the surface has the scaled strain
  rate lower than 0.1 (shown by dotted in (c)). (d) Nusselt number and
  (e) root-mean-square surface velocity as a function of time (shown
  in gray). For this model run, statistically steady state was
  achieved at $t^* = 1.8$, and running average is taken from the
  subsequent model results. In (d) and (e), the running average and
  its uncertainty (1 $\sigma$) is shown as solid and dotted lines,
  respectively, and the one standard deviation of the temporal
  variation itself is shown by dashed line.}
\label{fig:runex}
\end{figure*}

A typical snapshot of model run is shown in Figure~\ref{fig:runex}a.
This is the case of $\gamma$=0.6, $\theta_0$=15, and $H^*$=20, with the
standard pseudoplastic rheology [equation~(\ref{eq:visc0})]. In
addition to $T^*_{\mbox{\scriptsize max}}$, I calculate two more
measures for the temperature scale. One is the domain-average temperature,
\begin{linenomath*}\begin{equation}
\langle T^* \rangle = \int \!\!\!\!\int T^* dx^* dz^* 
\left/ \int\!\!\!\!\int dx^* dz^* \right.,
\end{equation}\end{linenomath*}
and the other is the (self-consistent) internal temperature \cite[]{korenaga09a},
\begin{linenomath*}\begin{equation}
T^*_i = \frac{1}{1-\delta'} \int_{\delta'}^1 
\left( \int T^* dx^* \left/ \int dx^* \right)\right. dz^*,
\end{equation}\end{linenomath*}
where  $\delta' = T^*_i/H^*$. 

The vigor of convection can be quantified by calculating the
root-mean-square velocity $v^*_{\mbox{\scriptsize rms}}$, and a
velocity diagnostic most indicative of the mode of convection is the
root-mean-square surface velocity $v^*_s$. To quantify how plate-like
the surface velocity field is, \cite{weinstein92} introduced the
notion of `plateness', and for convection exhibiting multiple plates
with different velocities, I use the following definition of
plateness,
\begin{linenomath*}\begin{equation}
P_{x} = \int_{e'<x} dx^* \left/ \int dx^* \right.,
\end{equation}\end{linenomath*}
where
\begin{linenomath*}\begin{equation}
e' = \frac{1}{v^*_s} \left|\frac{dv^*(z^*=0)}{dx^*}\right|.
\end{equation}\end{linenomath*}
The parameter $P_x$ measures the fraction of surface with normalized
strain rate $e'$ smaller than the given threshold $x$. The velocity
profile shown in Figure~\ref{fig:runex}b, for example, has $P_{0.1}$
of 0.76 (Figure~\ref{fig:runex}c). Like other definitions of
plateness, $P_x$ varies from 0 to 1, with higher values corresponding
to more rigid behavior. For comparison, actual plates on Earth tend to
have wide diffuse boundary zones, which occupy $\sim$15\% of the
surface area at the present day \cite[]{gordon92}.

To understand the spatial distribution of viscous dissipation,
I define $\Phi_c$ as the viscous dissipation within the region above
$z^*=c$, 
\begin{linenomath*}\begin{equation}
\Phi_c = \int_0^c \left( \int \eta^* e^*_{ij} e^*_{ij} dx^* \right) dz^*, 
\end{equation}\end{linenomath*}
and calculate $\Phi_{\delta}$, $\Phi_{\delta/2}$, and $\Phi$ ($\equiv
\Phi_1$), where $\delta$ is (on average) the maximum thickness of the
top thermal boundary layer \cite[e.g.,][]{busse67a},
\begin{linenomath*}\begin{equation}
\delta = 2 Nu^{-1}.
\label{eq:delta}
\end{equation}\end{linenomath*} 
Here $\delta$ is nondimensionalized by the model depth $D$. 

Each case was run up to $t^*=6$. The cumulative heat generation and
cumulative heat loss from the surface are monitored, and when these
two start to match within $\sim$1\%, I judge that the system has
reached statistically steady state. This usually takes place at $t^* =
\sim2-3$, and I use subsequent model results to calculate time-average
values of key diagnostics such as $Nu$ and $\Phi$. The (one) standard
deviation of time-averaged $Nu$ is typically less than 1\%, whereas
the standard deviation of $Nu$ itself is often greater by one order of
magnitude, reflecting the highly time-dependent nature of convection
(Figure~\ref{fig:runex}d). Surface velocity exhibits even greater
time dependency (Figure~\ref{fig:runex}e).

\subsection{Reference Scaling}

A total of 82 cases were run with the standard pseudoplastic rheology
using different combinations of $\gamma$ (0.1-1), $\theta_0$ (10-25),
and $H^*$ (8-20). The summary of convection diagnostics is reported in
Tables~\ref{table:runKa} and \ref{table:runKb}; eight runs resulted in
stagnant-lid convection, and others exhibited plate-tectonic
convection.

\begin{figure*}[t]
\small
\singlespacing
\begin{center}
\epsfig{file=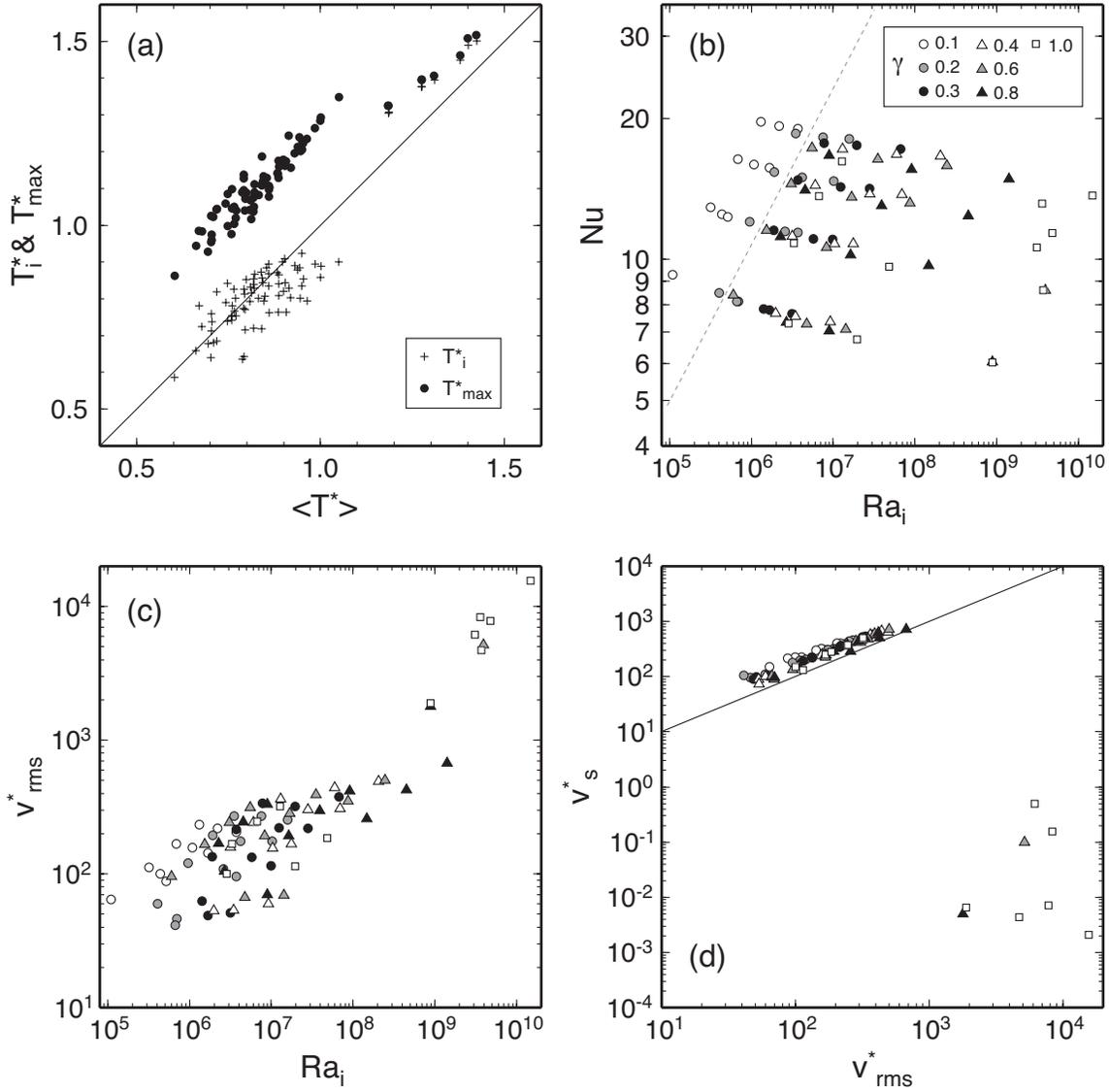,height=15cm}
\end{center}
\caption{Correlations among convection diagnostics from
  reference runs.  (a) Internal temperature $T^*_i$ (cross) and
  maximum temperature $T^*_{\mbox{\scriptsize max}}$ (solid circle)
  are compared with domain-average temperature $\langle T^*
  \rangle$. (b) $Nu$ and $Ra_i$. Dashed line indicates the slope of
  $Ra_i^{1/3}$. (c) $v^*_{\mbox{\scriptsize rms}}$ and $Ra_i$. (d)
  $v^*_s$ and $v^*_{\mbox{\scriptsize rms}}$.  In (b)-(d), different
  symbols denote runs with different $\gamma$, as shown by the legend
  in (b). }
\label{fig:refraw}
\end{figure*}

Three different temperature scales, $T^*_{\mbox{\scriptsize max}}$,
$\langle T^* \rangle$, and $T^*_i$, are correlated well to each other
(Figure~\ref{fig:refraw}a). Regardless of the mode of convection,
$T^*_{\mbox{\scriptsize max}}$ is distinctly higher than $\langle T^*
\rangle$, and this is because the thickness of the top thermal
boundary layer is not trivial in those runs. The maximum $Nu$ achieved
is only $\sim$20 (Figure~\ref{fig:refraw}b), and these different
temperature scales are expected to converge as $Nu$ increases.
Stagnant-lid runs are characterized by similar $T^*_{\mbox{\scriptsize
    max}}$ and $T^*_i$, because of small temperature variations
beneath the stagnant lid.

The relation between $Nu$ and $Ra_i$ appears to roughly follow the
classical scaling of $Nu \propto Ra_i^{1/3}$ within runs with the same
$\gamma$ and similar $\theta$ (Figure~\ref{fig:refraw}b). Varying
$\gamma$ has considerable effects on the scaling of $Nu$ as well as
$v^*_{\mbox{\scriptsize rms}}$ (Figure~\ref{fig:refraw}c).  The
distinction between plate-tectonic and stagnant-lid runs is very clear
in the correlation (or lack thereof) between $v^*_{\mbox{\scriptsize
    rms}}$ and $v^*_s$ (Figure~\ref{fig:refraw}). Average surface
velocity in these plate-tectonic runs is higher than corresponding
$v^*_{\mbox{\scriptsize rms}}$ because the latter involves averaging
over the entire domain, the majority of which moves more slowly than
surface plates.

My scaling analysis to understand the systematics of these model
results is based on the local stability of top thermal boundary layer
\cite[]{howard66}. Because of pseudoplastic rheology, the effective
viscosity of the top boundary layer or the effective lithospheric
viscosity, $\eta_L$, is expected to be higher than the interior
viscosity, and I denote the viscosity contrast between them as
\begin{linenomath*}\begin{equation}
\Delta \eta_L = \eta_L / \eta_i.
\end{equation}\end{linenomath*}
Viscosity in the top thermal boundary layer varies considerably as
specified by equation~(\ref{eq:visc0}), and the effective lithospheric
viscosity is an attempt to capture the overall stiffness of the
boundary layer by just one viscosity value. For the stiff boundary
layer to subduct, it has to become convectively unstable at least, and
by assuming that the maximum thickness of the boundary layer $\delta\,
D$ corresponds to marginal stability, the following relation should
hold:
\begin{linenomath*}\begin{equation}
\frac{\alpha \rho_0 g (T^*_{\mbox{\scriptsize max}} \Delta T) (\delta D)^3}{\kappa \eta_L}
= Ra_c,
\end{equation}\end{linenomath*}
where $Ra_c$ is the critical Rayleigh number. By using the relation
between $\delta$ and $Nu$ [equation~(\ref{eq:delta})], this marginal
stability criterion may be rearranged as
\begin{linenomath*}\begin{equation}
Nu = 2 \left( \frac{Ra_i}{Ra_c} \right)^{1/3} \Delta \eta_L^{-1/3},
\label{eq:Nu}
\end{equation}\end{linenomath*}
or
\begin{linenomath*}\begin{equation}
\Delta \eta_L = \frac{8 Ra_i}{Ra_c \, Nu^3}.
\label{eq:viscL}
\end{equation}\end{linenomath*}
Hereinafter $Ra_c$ is set to $10^3$. 

\begin{figure*}[t]
\small
\singlespacing
\begin{center}
\epsfig{file=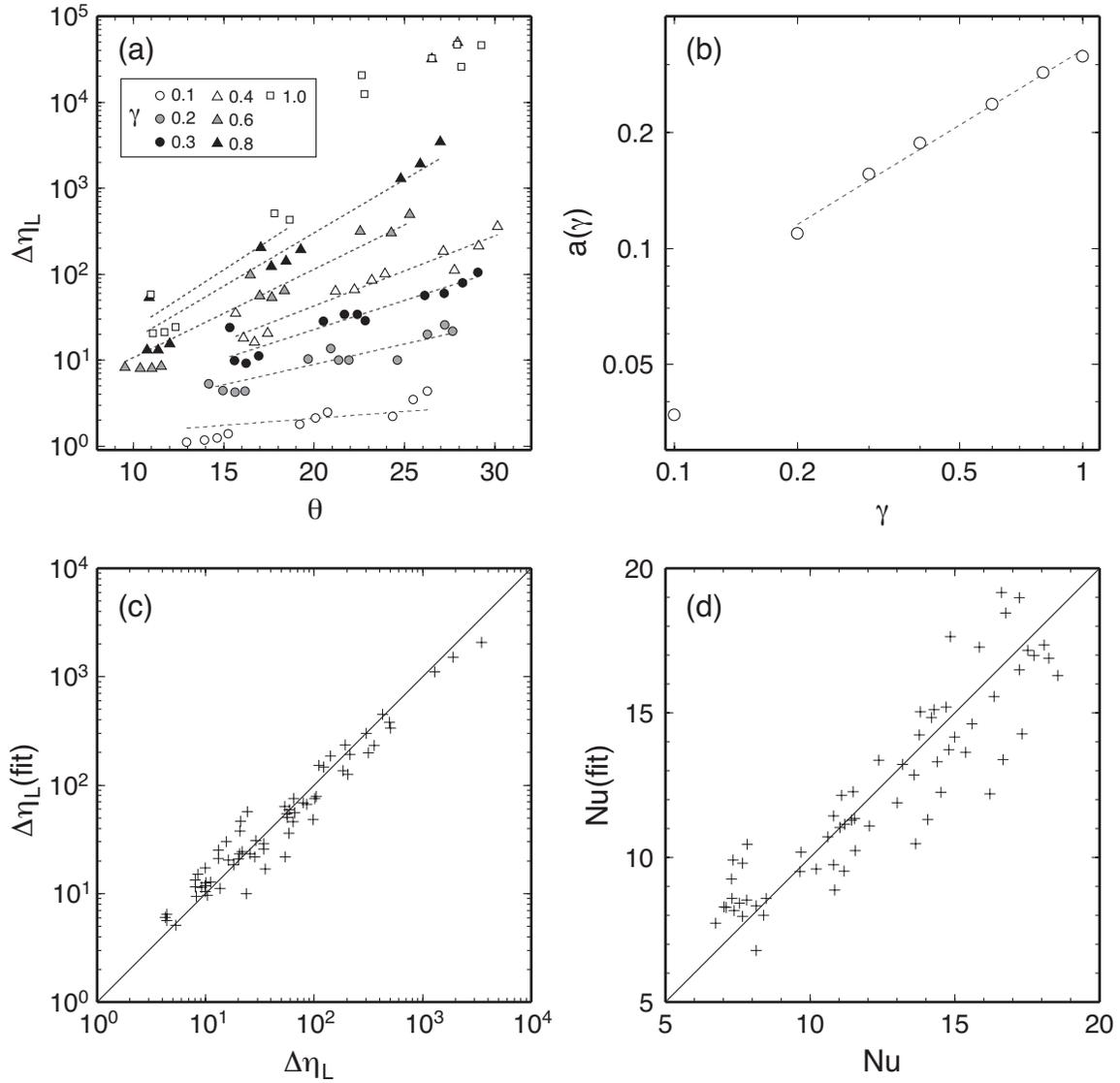,height=15cm}
\end{center}
\caption{The systematics of reference runs can be summarized
  through effective lithospheric viscosity contrast $\Delta \eta_L$.
  (a) $\Delta \eta_L$ as a function of $\theta$. Different symbols
  denote runs with different $\gamma$, and dashed lines are fitted
  trend in the form of equation~(\ref{eq:viscLfunc}) for each
  $\gamma$.  Data with $\Delta \eta_L$ greater than $10^4$ are
  stagnant-lid runs, which are excluded from linear regression. (b)
  The fitted coefficient $a$ as a function of $\gamma$. Dashed line
  represents equation~(\ref{eq:agamma}). (c) Comparison of measured
  $\Delta \eta_L$ with predicted values based on equations~(\ref{eq:viscLfunc})
  and (\ref{eq:agamma}). (d) Comparison of measured $Nu$ with predicted values
  based on equation~(\ref{eq:Nu}).}
\label{fig:refviscL}
\end{figure*}

Equation~(\ref{eq:viscL}) may be regarded as a way to extract
lithospheric viscosity contrasts from the measured pairs of $Ra_i$ and
$Nu$. The lithospheric viscosity contrast calculated this way
increases as $\theta$ increases, and this $\theta$ sensitivity is
greater for higher $\gamma$ (Figure~\ref{fig:refviscL}a).  The
following functionality appears to be sufficient to reproduce the
first-order behavior of the viscosity contrast,
\begin{linenomath*}\begin{equation}
\Delta \eta_L (\gamma,\theta)= \exp[a(\gamma) \theta],
\label{eq:viscLfunc}
\end{equation}\end{linenomath*}
which converges to unity at the limit of zero $\theta$.  The
coefficient $a(\gamma)$ is determined by linear regression for each
group of runs with the same $\gamma$ (Figure~\ref{fig:refviscL}a).
Excluding the result for $\gamma$ of 0.1, the runs with which are
characterized by rather low plateness (Table~\ref{table:runKa}), the
coefficient is linearly correlated with $\gamma$ in the logarithmic
space (Figure~\ref{fig:refviscL}b), which may be expressed as
\begin{linenomath*}\begin{equation}
a(\gamma) \approx 0.327 \gamma^{0.647}. 
\label{eq:agamma}
\end{equation}\end{linenomath*}
Equations~(\ref{eq:viscLfunc}) and (\ref{eq:agamma}) can predict
$\Delta \eta_L$ reasonably well over the range of four orders of
magnitude (Figure~\ref{fig:refviscL}c). The prediction for $Nu$
through equation~(\ref{eq:Nu}) has the average error of $\sim$10\%
(Figure~\ref{fig:refviscL}d). This error is considerably larger than
that observed for the heat-flow scaling derived by \cite{moresi98},
which was based on convection with the aspect ratio of one. The use of
the wide aspect ratio ($=8$) in this study and resultant
time dependency in convection patterns may be the source of these
scatters.

\begin{figure*}[t]
\small
\singlespacing
\begin{center}
\epsfig{file=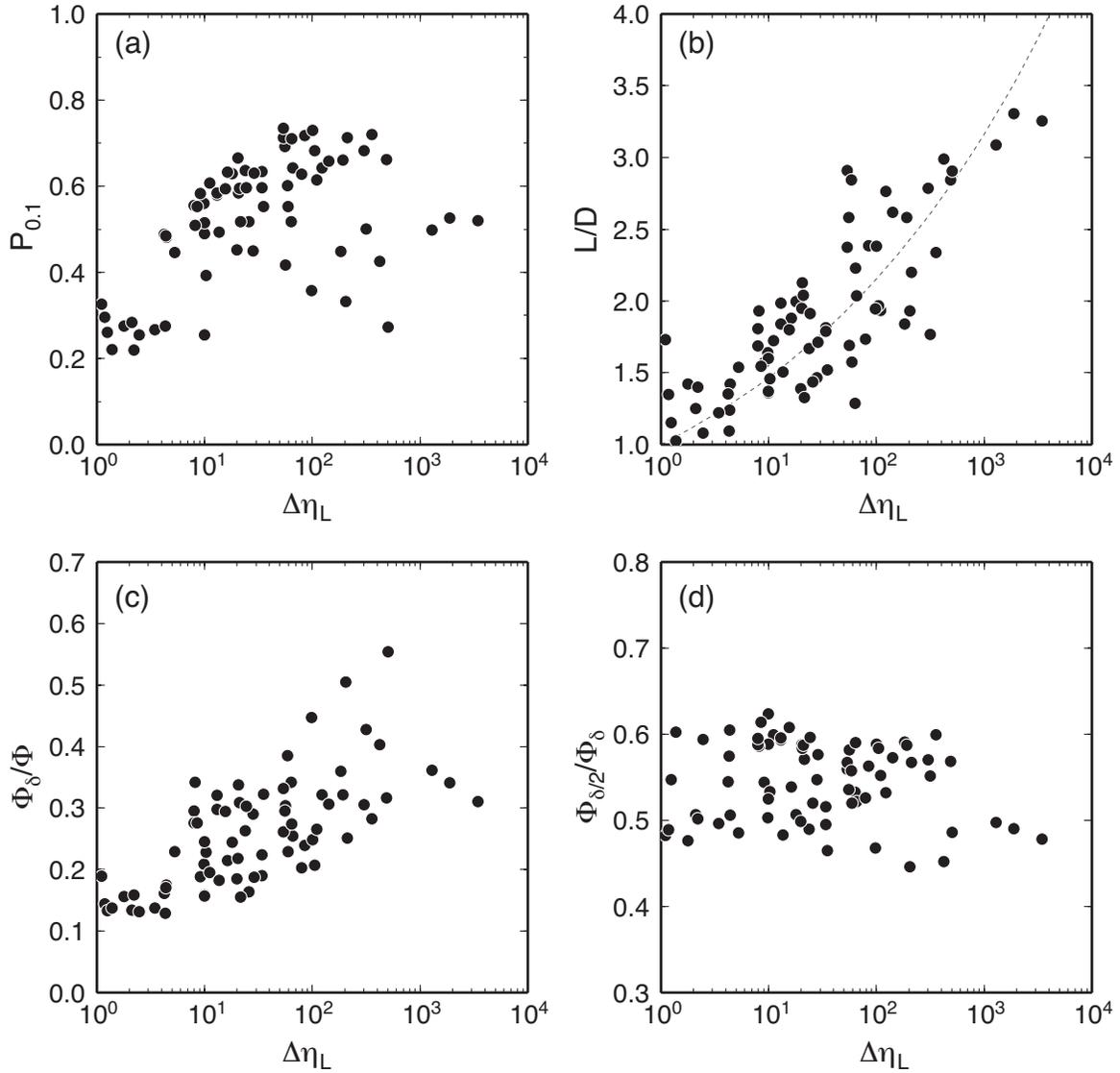,height=15cm}
\end{center}
\caption{Covariations of measured lithospheric viscosity
  contrast $\Delta \eta_L$ with (a) plateness $P_{0.1}$, (b) the
  average aspect ratio of convection cells $L/D$, (c) the fraction of
  viscous dissipation taking place in the top thermal boundary layer
  with respect to that in the entire domain $\Phi_{\delta}/\Phi$, and
  (d) the fraction of viscous dissipation in the upper half of the
  boundary layer with respect to that in the entire boundary layer
  $\Phi_{\delta/2}/\Phi_{\delta}$. Dashed line in (b) corresponds to
  $L/D = \Delta \eta_L^{1/6}$. }
\label{fig:refcorr}
\end{figure*}

The lithospheric viscosity contrast $\Delta \eta_L$ calculated from
equation~(\ref{eq:viscL}) exhibits broad correlations with other
convection diagnostics (Figure~\ref{fig:refcorr}). Higher $\Delta
\eta_L$ generally gives rise to higher plateness
(Figure~\ref{fig:refcorr}a) and greater viscous dissipation in the top
boundary layer (Figure~\ref{fig:refcorr}c).  How viscous dissipation
is distributed within the boundary layer, however, seems to be
insensitive to variations in $\Delta \eta_L$ as
$\Phi_{\delta/2}/\Phi_{\delta} \sim$ 0.5-0.6 for all of plate-tectonic
runs (Figure~\ref{fig:refcorr}d). 

Assuming the half-space cooling of lithosphere, the (average) maximum
plate thickness $\delta$ is related to the average length of plates,
$L$, as
\begin{linenomath*}\begin{equation}
\delta \sim 2 \left( \frac{L}{D v^*_s} \right)^{1/2},
\end{equation}\end{linenomath*}
where $L/(D v^*_s)$ is the average time from a ridge to a subduction
zone. Thus, the average aspect ratio of convection cells may be
calculated from $Nu$ and $v^*_s$ as
\begin{linenomath*}\begin{equation}
\frac{L}{D} = \frac{v^*_s}{Nu^2},
\label{eq:LD}
\end{equation}\end{linenomath*}
for which equation~(\ref{eq:delta}) is used. 
The aspect ratio gradually increases as $\Delta \eta_L$ increases,
i.e., stronger plates tend to be longer (Figure~\ref{fig:refcorr}b).
By assuming some empirical relation for $L/D$ (e.g., $L/D \sim \Delta
\eta_L^{1/6}$), equation~(\ref{eq:LD}) may be rearranged as scaling
for $v^*_s$,
\begin{linenomath*}\begin{equation}
v^*_s = 4 \left(\frac{L}{D}\right) \left(\frac{Ra_i}{Ra_c}\right)^{2/3}
\Delta \eta_L^{-2/3}.
\end{equation}\end{linenomath*}
That is, unlike the scaling for heat flux [equation~(\ref{eq:Nu})],
some information on the aspect ratio of convection cells is essential
for the scaling for surface velocity.

\subsection{Effect of Shallow Stiffening}

On Earth, the creation of new plates at mid-ocean ridges is usually
accompanied by the melting of upwelling mantle (unless the mantle is
too cold), and this chemical differentiation along the global
mid-ocean ridge system constitutes the dominant fraction of
terrestrial magmatism \cite[]{crisp84}. This mantle melting results in
the formation of oceanic crust as well as depleted mantle lithosphere,
both of which are chemically more buoyant with respect to the
underlying asthenosphere \cite[]{oxburgh77}, and the depleted
lithosphere also becomes intrinsically more viscous (by $\sim$$10^3$)
because of dehydration caused by melting \cite[]{hirth96}. As long as
plate tectonics is taking place, the chemical buoyancy of oceanic
lithosphere is insignificant as resistance to subduction because the
basalt-to-eclogite transition at relatively shallow depth ($<$60~km)
makes the subducting slab compositionally denser than the surrounding
mantle \cite[e.g.,][]{ringwood88}. In this study, therefore, I focus
on the effect of dehydration stiffening on the scaling of
plate-tectonic convection. As in \cite{korenaga09a}, instead of
tracing the advection of the dehydrated slab through time, I use the
depth-dependent viscosity that is fixed in time
[equation~(\ref{eq:zdep})].  Though the subducting slab loses the
extra viscosity contrast $\Delta \eta$ as soon as it passes the given
depth $h^*$, the effect of shallow stiffening on slab bending can
still be evaluated with this scheme.

The scaling of $Nu$ [equation~(\ref{eq:Nu})] indicates that different
combinations of $Ra_i$ and $\Delta \eta_L$ can produce the same
$Nu$. Thus, in terms of the efficiency of heat transport, a run with
high $Ra_i$ and high $\Delta \eta_L$ may be indistinguishable from
that with low $Ra_i$ and low $\Delta \eta_L$, but shallow stiffening
may affect these cases differently. A wide variety of plate-tectonic
cases were thus simulated by varying $\gamma$ (0.4-0.8), $\theta_0$
(10-25), and $H^*$ (2-20), and shallow stiffening was incorporated
with $h^*$ ranging from 0.1 to 0.3 and $\Delta \eta$ ranging from 3 to
$10^3$. Stiffening by mantle melting is limited mostly to the top
200~km or so (i.e., $h^* < 0.07$), and the use of greater $h^*$ is to
study the asymptotic behavior of stiffening effects. Also, the top
thermal boundary layer in numerical modeling is thicker than actual
oceanic lithosphere because of relatively low $Ra_i$ used in this
study, so $h^*$ has to be comparably large in order to reproduce an
Earth-like combination of thermal and compositional boundary layers.
The number of runs is 225 in total, with 19 stagnant-lid runs
(Tables~\ref{table:runLa1}-\ref{table:runLe}).  Convection diagnostics
were measured in the same way for reference runs.

\begin{figure*}[t]
\small
\singlespacing
\begin{center}
\epsfig{file=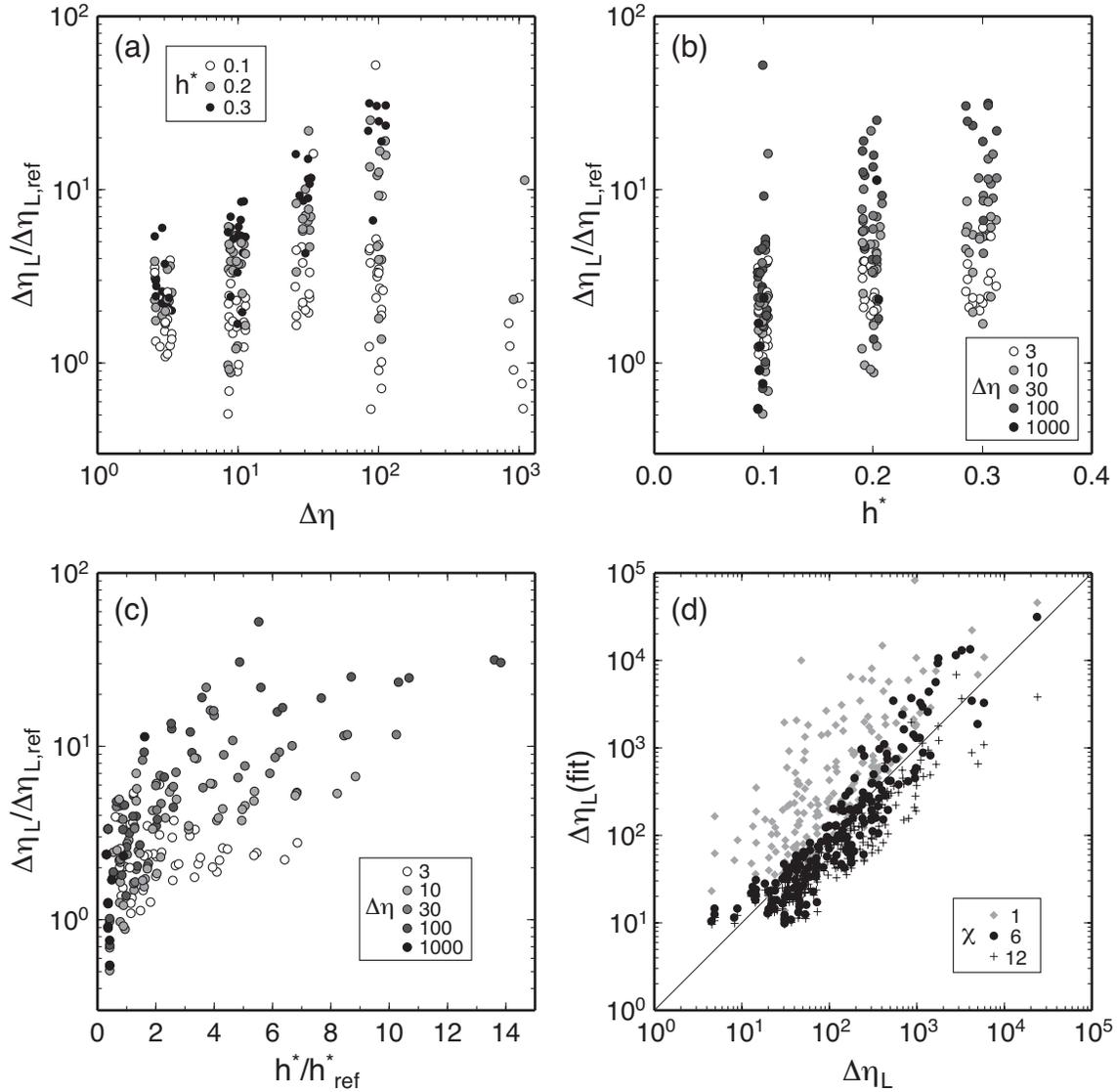,height=15cm}
\end{center}
\caption{The effect of additional depth-dependent viscosity on
  the lithospheric viscosity contrast. The deviation from the
  prediction based on standard pseudoplastic rheology, $\Delta
  \eta_L/\Delta \eta_{L,\mbox{\scriptsize ref}}$, is shown as a
  function of (a) the viscosity contrast due to dehydration $\Delta
  \eta$, (b) the thickness of dehydrated layer $h^*$, and (c) the
  same thickness but scaled by the reference thickness,
  $h^*/h^*_{\mbox{\scriptsize ref}}$. Different symbols denote
  different $h^*$ in (a) and different $\Delta \eta$ in (b) and
  (c). Note that in (a) and (b) the values of $\Delta \eta$ and $h^*$
  are slightly perturbed randomly for display purposes. (d) Comparison
  of measured $\Delta \eta_L$ with predicted values based on
  equation~(\ref{eq:dviscLchi}), for three different values of
  $\chi$. }
\label{fig:depthviscL}
\end{figure*}

The scaling analysis for runs with shallow stiffening is founded on
that for the reference runs. As in the previous section,
the lithospheric viscosity contrast $\Delta \eta_L$ is calculated from
the measured pair of $Nu$ and $Ra_i$. A key issue is how this
viscosity contrast is influenced by the additional depth-dependent
viscosity, and this influence may be measured by the deviation from
the prediction based on equation~(\ref{eq:viscLfunc}).  The predicted
viscosity contrast is based solely on $\gamma$ and $\theta$, and it is
denoted as $\Delta \eta_{L,\mbox{\scriptsize ref}}$ to distinguish
from the actual $\Delta \eta_L$.  The ratio $\Delta \eta_L/\Delta
\eta_{L,\mbox{\scriptsize ref}}$ is loosely correlated with $\Delta
\eta$ and $h^*$ as one may expect (Figure~\ref{fig:depthviscL}a,b);
higher $\Delta \eta$ or $h^*$ leads to higher $\Delta \eta_L$ than
predicted by equation~(\ref{eq:viscLfunc}). A better correlation may
be seen between the two ratios, $\Delta \eta_L/\Delta
\eta_{L,\mbox{\scriptsize ref}}$ and $h^*/h^*_{\mbox{\scriptsize
    ref}}$ (Figure~\ref{fig:depthviscL}c), where
$h^*_{\mbox{\scriptsize ref}}$ is defined as
\begin{linenomath*}\begin{equation}
h^*_{\mbox{\scriptsize ref}} = Nu_{\mbox{\scriptsize ref}}^{-1},
\label{eq:href}
\end{equation}\end{linenomath*}
and
\begin{linenomath*}\begin{equation}
Nu_{\mbox{\scriptsize ref}} = 2 \left( \frac{Ra_i}{Ra_c} \right)^{1/3} 
\Delta \eta_{L,\mbox{\scriptsize ref}}^{-1/3}.
\end{equation}\end{linenomath*}
The parameter $h^*_{\mbox{\scriptsize ref}}$ is the averaged thickness
of the top thermal boundary layer expected for a run with the same
$Ra_i$ but with the standard pseudoplastic rheology. The ratio $\Delta
\eta_L / \Delta \eta_{L,\mbox{\scriptsize ref}}$ increases as
$h^*/h^*_{\mbox{\scriptsize ref}}$ but eventually saturates and never
exceeds the given $\Delta \eta$. This behavior may be represented by
the following functionality,
\begin{linenomath*}\begin{equation}
\Delta \eta_L = \Delta \eta_{L,\mbox{\scriptsize ref}} \,
\exp\left[ \ln(\Delta \eta) \max\left(1,
\frac{h^*}{\chi h^*_{\mbox{\scriptsize ref}}}\right)\right],
\label{eq:dviscLchi}
\end{equation}\end{linenomath*}
which means that $\Delta \eta_L$ converges to the simple product of
$\Delta \eta_{L,\mbox{\scriptsize ref}}$ and $\Delta \eta$ when $h^*$
is sufficiently greater than $h^*_{\mbox{\scriptsize ref}}$.  The
parameter $\chi$ controls how high $h^*$ should be with respect to
$h^*_{\mbox{\scriptsize ref}}$ in order to achieve the convergence;
greater $\chi$ means that thicker $h^*$ is required. By trying a range
of values, I found that $\chi \approx 6$ can reproduce the measured
$\Delta \eta_L$ reasonably well (Figure~\ref{fig:depthviscL}d).

\begin{figure*}[t]
\small
\singlespacing
\begin{center}
\epsfig{file=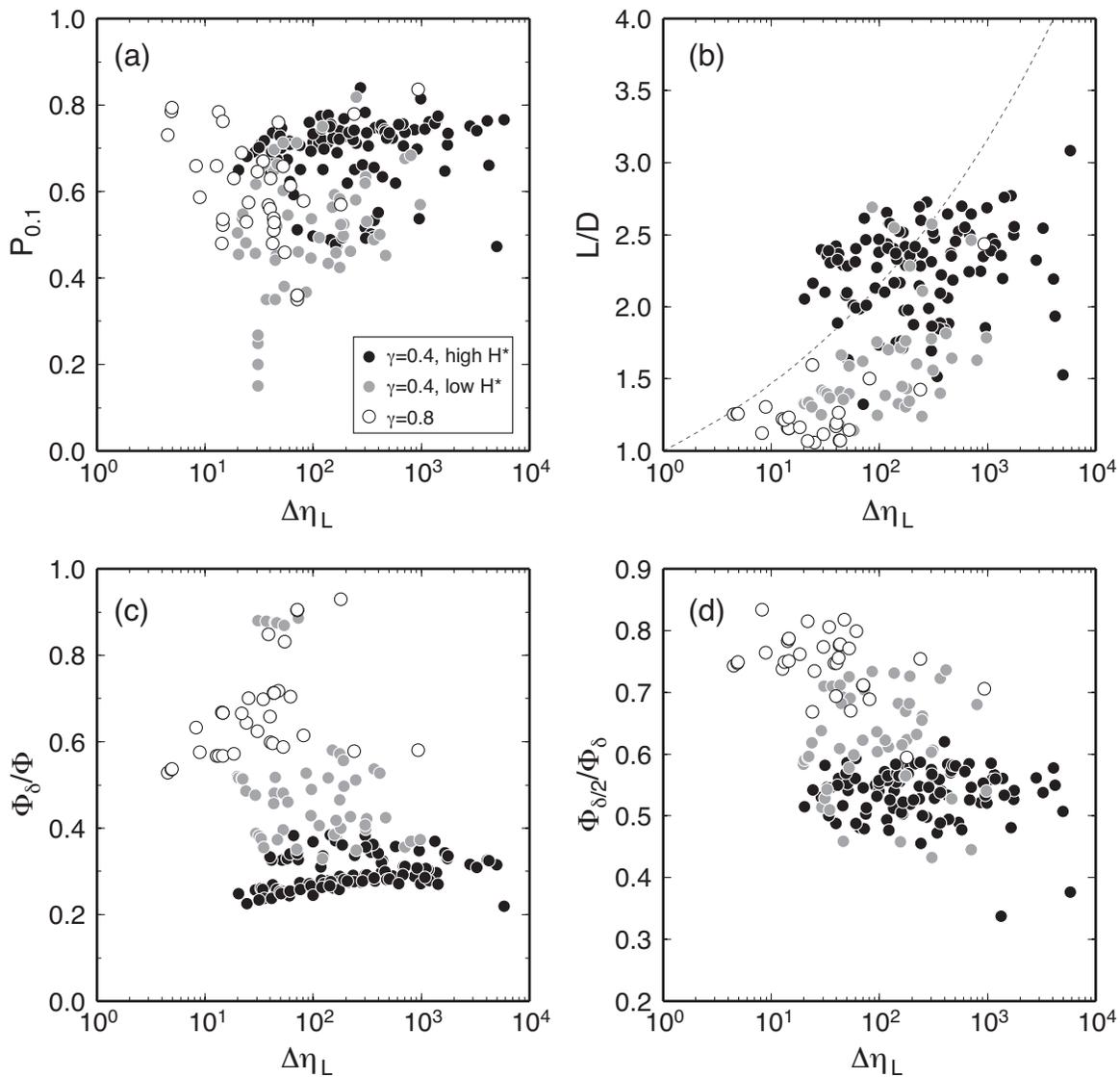,height=15cm}
\end{center}
\caption{Same as Figure~\ref{fig:refcorr} but for runs with
  shallow stiffening.  Different symbols correspond to different
  groups of runs: $\gamma$ of 0.4 with high $H^*$ (solid circle;
  Tables~\ref{table:runLa1}-\ref{table:runLa3}), $\gamma$ of 0.4 with
  low $H^*$ (gray circle;
  Tables~\ref{table:runLb1}-\ref{table:runLb2}), and $\gamma$ of 0.8
  (open circle; Table~\ref{table:runLe}). }
\label{fig:depthcorr}
\end{figure*}

The relations between $\Delta \eta_L$ and other convection diagnostics
are more ambiguous than observed for the reference runs
(Figure~\ref{fig:depthcorr}). It appears to be premature to
parameterize the aspect ratio $L/D$ as a simple function of the
lithospheric viscosity contrast (Figure~\ref{fig:depthcorr}b), and
more thorough work is clearly required to better understand the
self-organization of plate tectonics. It is still interesting to note,
however, that high plateness is possible even with low viscosity
contrast (Figure~\ref{fig:depthcorr}a) and that plate-tectonic
convection can occur even when most of viscous dissipation takes place
in the top boundary layer (Figure~\ref{fig:depthcorr}c).

\subsection{Conditions for Plate-Tectonic Convection}

\begin{figure*}[t]
\small
\singlespacing
\begin{center}
\epsfig{file=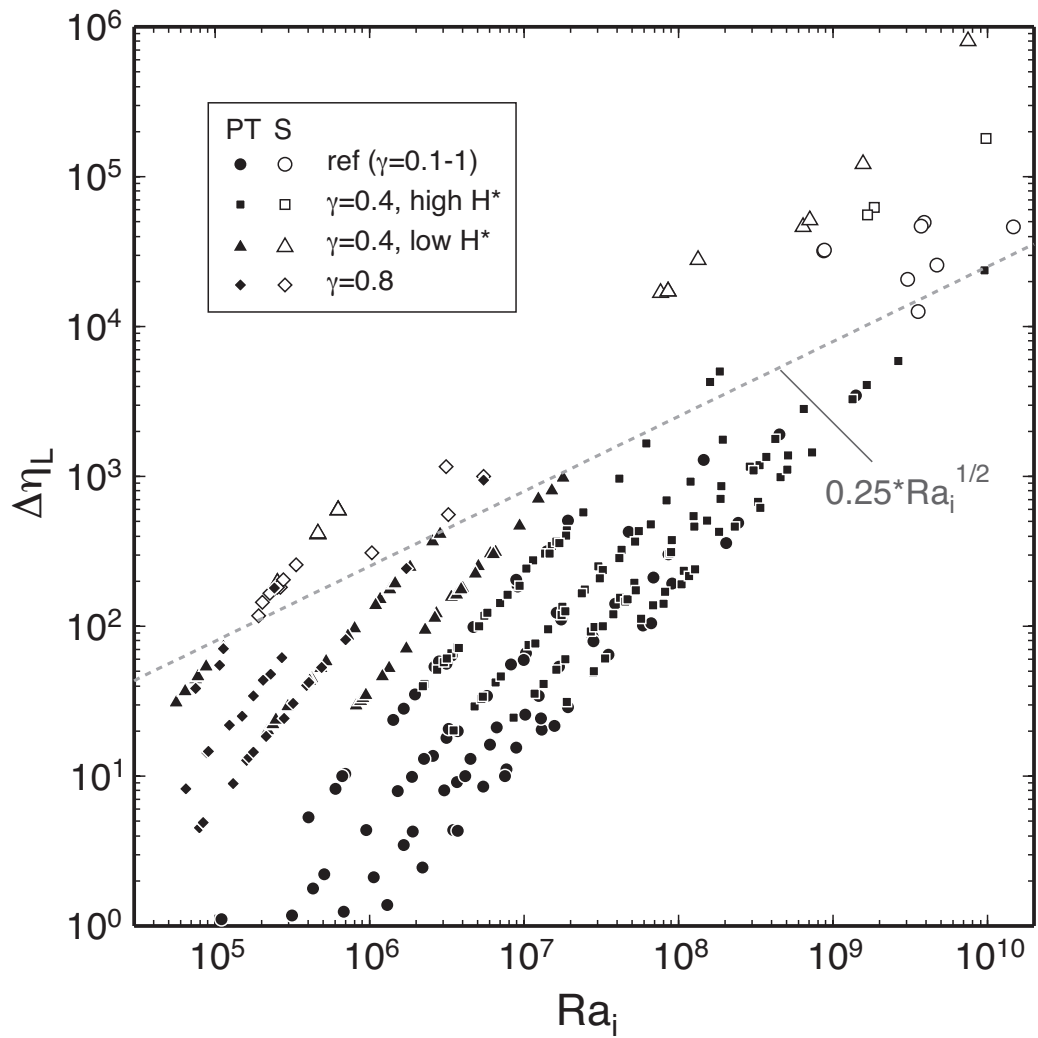,height=14cm}
\end{center}
\caption{Covariation of $Ra_i$ and $\Delta \eta_L$ for all
  model runs. Solid and open symbols denote plate-tectonic and
  stagnant-lid runs, respectively. Dashed line represents an
  approximate divide between these two modes of convection ($\Delta
  \eta_L \sim 0.25 Ra_i^{1/2}$).}
\label{fig:raivis}
\end{figure*}

The condition for plate-tectonic convection is found to be seen most
clearly in the covariation of $Ra_i$ and $\Delta \eta_L$
(Figure~\ref{fig:raivis}). Plate-tectonic convection is possible even
with high lithospheric viscosity contrast if $Ra_i$ is sufficiently
high, and the critical viscosity contrast, above which plate-tectonic
convection is unlikely, appears to be
\begin{linenomath*}\begin{equation}
\Delta \eta_{L,\mbox{\scriptsize crit}} \approx 0.25 Ra_i^{1/2},
\end{equation}\end{linenomath*}
though there are a few runs that slightly violate this threshold.

To understand the meaning of this scaling, I consider stress balance
at the bending of subducting slab, which is probably the most critical
part of plate-tectonic convection. First, the stress due to the
negative buoyancy of the slab may be expressed as
\begin{linenomath*}\begin{equation}
\tau_S \sim \alpha \rho_0 g (T^*_{\mbox{\scriptsize max}} \Delta T) D,
\end{equation}\end{linenomath*}
or by normalizing the (internal) stress scale, $\eta_i \kappa/D^2$, 
\begin{linenomath*}\begin{equation}
\tau^*_S \sim Ra_i.
\end{equation}\end{linenomath*}
Second, the bending stress should be proportional to the lithospheric viscosity
and bending strain rate as \cite[e.g.,][]{conrad99a}
\begin{linenomath*}\begin{equation}
\tau_B \sim \eta_L \frac{v_s (\delta D)}{R^2},
\end{equation}\end{linenomath*}
where $R$ is the radius of curvature, and its nondimensionalized form
is 
\begin{linenomath*}\begin{equation}
\tau^*_B \sim \Delta \eta_L \, v^*_s \, \delta \left(\frac{D}{R}\right)^2
\propto \Delta \eta_L^{2/3} Ra_i^{1/3} \left(\frac{D}{R}\right)^2. 
\end{equation}\end{linenomath*}
Finally, by assuming $\tau_S \approx \tau_B$ at $\Delta \eta_L =
\Delta \eta_{L,\mbox{\scriptsize crit}}$, we may derive the following
scaling for the radius of curvature,
\begin{linenomath*}\begin{equation}
\frac{R}{D} \propto Ra_i^{-1/6}. 
\end{equation}\end{linenomath*}
Thus, the radius is weakly dependent of the vigor of convection, and
it becomes smaller for more vigorous convection. Without this
variation in the radius of curvature, the critical viscosity contrast
would be more sensitive to a change in $Ra_i$ (i.e., proportional to
$Ra_i$ instead of $Ra_i^{1/2}$).

Note that all of 307 runs reported here are either strictly
plate-tectonic or stagnant-lid convection, and there is no case of
episodic overturn, in which the system periodically goes back and
forth between plate-tectonic and stagnant-lid modes
\cite[]{moresi98}. This is consistent with the use of virtually zero
cohesion strength and finite friction coefficient in this study
(\S\ref{sec:rheology}), because the possibility of the episodic
overturn mode appears to be important only with nontrivial cohesion
strength \cite[e.g.,][]{moresi98,stein04,oneill07b}. The use of purely
internal heating (thus the lack of upwelling plumes) in this study
might also be responsible. 

\section{Discussion and Conclusion \label{sec:discussion}}

On the basis of the scaling of $Nu$ [equation~(\ref{eq:Nu})] and the
parameterization of $\Delta \eta_L$ [equations~(\ref{eq:viscLfunc}),
(\ref{eq:agamma}), and (\ref{eq:dviscLchi})], it is now possible to
discuss the relation between mantle temperature and surface heat flux,
which is fundamental to our theoretical understanding of the long-term
evolution of Earth. Because some of key model parameters are still
poorly known \cite[e.g.,][]{korenaga08a}, the following exercise
should be regarded as a preliminary case study. As explained below,
the self-consistent construction of a plausible heat-flow scaling law
requires modeling the thermal and chemical evolution of Earth at the
same time, so a more extensive exploration of the scaling of plate
tectonics will be reported elsewhere.

First, for the dependency of viscosity on mantle potential temperature
$T_p$, the following Arrhenius form is used,
\begin{linenomath*}\begin{equation}
\eta_T(T_p) = \eta_r \exp\left( \frac{E}{RT_p} - \frac{E}{RT_r} \right),
\end{equation}\end{linenomath*}
where $\eta_r$ is reference viscosity at $T_p = T_r$, and $E$ is
assumed to be 300~kJ~mol$^{-1}$ \cite[]{korenaga06}. The reference
temperature is set to 1623~K (1350$^{\circ}$C), which corresponds to
the present-day potential temperature of the ambient mantle
\cite[]{herzberg07}. The Frank-Kamenetskii parameter $\theta$ is
calculated from equation~(\ref{eq:theta}), and the reference
lithospheric viscosity contrast $\Delta \eta_{L,\mbox{\scriptsize
    ref}}$ is calculated with $\gamma$ of 0.8, which corresponds to
the effective friction coefficient of $\sim$0.02
[equation~(\ref{eq:gammadef})].

\begin{figure*}[t]
\small
\singlespacing
\begin{center}
\epsfig{file=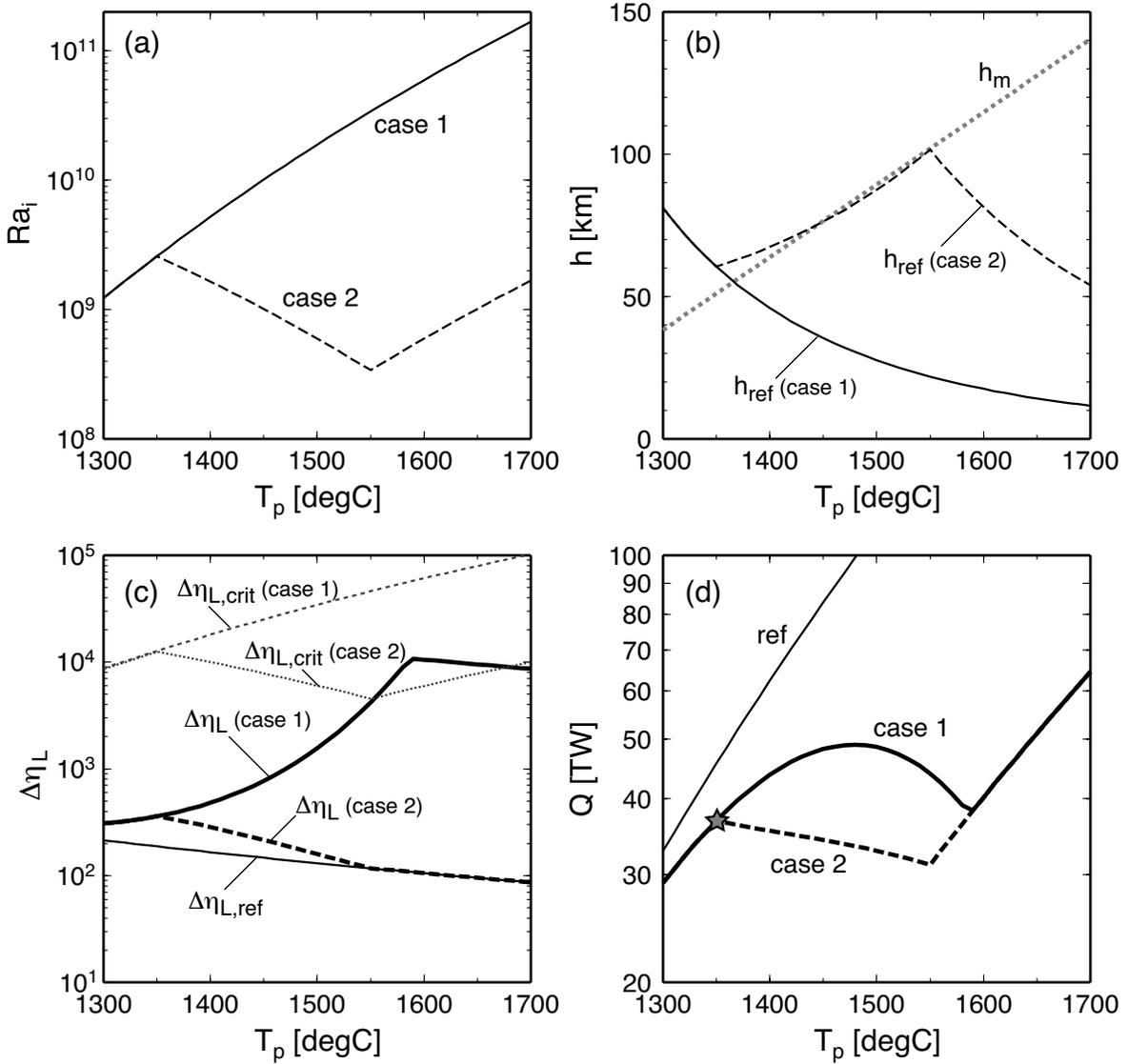,height=15cm}
\end{center}
\caption{A worked example of how the new scaling laws of
  plate-tectonic convection may be used to build heat-flow scaling for
  Earth. (a) Internal Rayleigh number $Ra_i$ as a function of mantle
  potential temperature $T_p$. (b) Thicknesses of dehydrated
  lithosphere ($h_m$, dotted) and reference thermal boundary layer
  ($h_{\mbox{\scriptsize ref}}$, solid for case~1 and dashed for
  case~2) [equation~(\ref{eq:href})]. (c) Lithospheric viscosity
  contrast for case~1 (thick solid), case~2 (thick dashed), and a
  reference case with no effect of mantle melting, i.e., $h^*_m=0$
  (thin solid). Also shown are the critical viscosity contrast for
  plate-tectonic convection (dashed for case~1 and dotted for
  case~2). (d) Relation between $T_p$ and surface heat flux
  $Q$. Legend is the same as in (c). Star denotes convective heat flux
  at the present day (38~TW at 1350$^{\circ}$C). See the main text for
  details.}
% note: why the solidus is at 50km for Tp=1350degC is just because I used
% rho=4000kg/m^3 when converting pressure to depth in the matlab script.
\label{fig:QvsT}
\end{figure*}

The internal Rayleigh number $Ra_i$ is then calculated with the above
temperature-viscosity and the following values: $\alpha$=$2 \times
10^{-5}$~K$^{-1}$, $\rho_0$=$4000$~kg~m$^{-3}$, $g$=$9.8$~m~s$^{-2}$,
and $D$=$2900 \times 10^3$~m. The dehydration of the mantle beneath
mid-ocean ridges is assumed to take place when the upwelling mantle
crosses the solidus for dry pyrolitic mantle, and the initial pressure
of melting (in GPa) can be calculated from the potential temperature
(in K) as \cite[]{korenaga02b}
\begin{linenomath*}\begin{equation}
P_o = (T_p-1423)/100,
\end{equation}\end{linenomath*}
and the thickness of dehydrated mantle $h_m$ is given by $P_o/(\rho_o
g)$. The nondimensional thickness $h^*$ is $h_m/D$, and the viscosity
contrast due to dehydration $\Delta \eta$ is assumed to be $10^2$
here.  The lithospheric viscosity contrast $\Delta \eta_L$ gradually
increases for higher $T_p$ (Figure~\ref{fig:QvsT}c, case~1) because hotter
mantle starts to melt deeper (Figure~\ref{fig:QvsT}b), but the effect
of shallow stiffening on the viscosity contrast saturates at $T_p
\sim$ 1600$^{\circ}$C, above which the contrast slightly decreases
because of smaller $\theta$ for higher $T_p$
[equation~(\ref{eq:theta})].

Lastly, the global heat flux $Q$ is calculated as
\begin{linenomath*}\begin{equation}
Q = k \, A \, T_p \, Nu / D,
\end{equation}\end{linenomath*}  
where $A$ is the surface area of Earth, and $k$ is assumed to be
4~W~m$^{-1}$~K$^{-1}$. The reference viscosity $\eta_r$ is set to
$10^{19}$~Pa~s so that the predicted global heat flux matches the
present-day convective heat flux of $\sim$38~TW \cite[]{korenaga08d}
(Figure~\ref{fig:QvsT}d, case~1). 
The effect of shallow stiffening suppresses
the heat flux considerably and even reverts the sense of temperature
sensitivity; the flux is lower for higher $T_p$ above
$\sim$1450$^{\circ}$C until the effect of shallow stiffening becomes
saturated at $\sim$1600$^{\circ}$C. For the temperature range of
1350-1600$^{\circ}$C, which is most relevant to the thermal evolution
of Earth for the last 3.5~Gyr \cite[]{herzberg10}, the predicted
relation between the mantle temperature and surface heat flux closely
resembles that suggested by \cite{korenaga06} along a similar line of
reasoning but on the basis of the global energy balance.

Note that the reference viscosity $\eta_r$ of $10^{19}$~Pa~s (at
present-day potential temperature) may be appropriate for
asthenosphere but would typically be regarded as too low to represent
the whole mantle, for which the viscosity of $10^{21}$-$10^{22}$~Pa~s
is usually assumed. The effective lithospheric viscosity contrast is
$\sim$3$\times 10^2$ in this example (Figure~\ref{fig:QvsT}c), which
is comparable to the discrepancy. Traditionally, the average viscosity
of the present-day mantle is estimated to be of that magnitude, in
order to explain surface heat flux (or equivalently, plate velocities)
\cite[e.g.,][]{hager91,bercovici00}, because the aforementioned geoid-based
studies (\S\ref{sec:strategy}) can constrain only relative variations
in viscosity and are insensitive to the absolute values of viscosity.
The heat-flow scaling of equation~(\ref{eq:Nu}) suggests that a
lithospheric viscosity contrast alone could regulate surface heat flux
without invoking a viscosity increase in the lower mantle. 

The use of constant viscosity contrast for dehydration stiffening
$\Delta \eta$ for the entire temperature range (thus implicitly over
the entire Earth history) is equivalent to assuming that the water
content of the convecting mantle does not change with time. If the
mantle is drier than present, for example, the viscosity contrast
would be smaller, and if the mantle is completely dry, mantle melting
should not cause any viscosity change. By combining the thermal budget
of Earth with geological constraints on sea level change and with the
growth of continental crust, \cite{korenaga08f} suggested that the
volume of Earth's oceans is unlikely to have been constant with time
and that the mantle may have been gradually hydrated by subduction
starting with a very dry state in the Archean.  As the second example
(denoted as case~2 in Figure~\ref{fig:QvsT}), I consider effective
heat-flow scaling expected for this scenario. For simplicity, the
mantle is assumed to have been hydrated linearly from the completely
dry state to the present state, as it cooled from 1550$^{\circ}$C, to
1350$^{\circ}$C for the last $\sim$3~Gyr \cite[]{herzberg10}. For
$T_p$ greater than 1550$^{\circ}$C, therefore the internal viscosity
is intrinsically higher by $\Delta \eta$, and this viscosity contrast
gradually diminishes as $T_p$ approaches 1350$^{\circ}$C. This is
reflected in how $Ra_i$ varies with $T_p$ (Figure~\ref{fig:QvsT}a,
case~2). At the same time, the viscosity contrast due to mantle
melting is unity at $T_p \ge $ 1550$^{\circ}$C and gradually increases
to the full value $\Delta \eta$ at $T_p$ of 1350$^{\circ}$C.  The
total lithospheric viscosity contrast $\Delta \eta_L$ in this scenario
is much reduced than the previous example (Figure~\ref{fig:QvsT}c),
but because of the overall reduction in $Ra_i$, the surface heat flux
is suppressed further, and the inverse relationship between mantle
temperature and heat flux dominates heat-flow scaling during the
mantle hydration period (Figure~\ref{fig:QvsT}d). Obviously, this type
of calculation should be done more self-consistently by modeling the
thermal evolution of Earth together with its global water cycle, and
what is presented here is only a crude estimate.

In both cases, the lithospheric viscosity contrast is always smaller
than its threshold (Figure~\ref{fig:QvsT}c), so plate-tectonic
convection seems to be dynamically plausible throughout the Earth
history, as long as surface water exists to hydrate the lithosphere
and reduce the effective friction coefficient \cite[]{korenaga07c}. In
particular, the gradually hydrating mantle (case~2) helps to maintain
relatively small $\Delta \eta_L$ even with deeper mantle melting at
higher $T_p$, facilitating the operation of plate tectonics in the
early Earth.

Though previous attempts to estimate the heat-flow scaling of plate
tectonics \cite[]{korenaga03a,korenaga06} have already predicted the
inverse relation between mantle temperature and surface heat flux as
indicated by Figure~\ref{fig:QvsT}, there are a few important
differences. First, because the effect of shallow stiffening
eventually saturates [equation~(\ref{eq:dviscLchi})], the inverse
relation is restricted to a certain temperature range. This subtle
behavior is difficult to derive from the global energy balance
approach adopted by the previous studies. Second, the global energy
balance can be exploited to derive heat-flow scaling by {\it assuming}
the mode of convection, so whether plate-tectonic convection is
plausible or not cannot be addressed. Finally, the energy balance
approach has a few poorly constrained parameters, such as the radius
of curvature for plate bending, effective lithospheric viscosity, and
the aspect ratio of convection, and it is possible to obtain wildly
different results by varying them independently
\cite[e.g.,][]{davies09a}. Though similarly suffering from parameter
uncertainty (e.g., $\gamma$ and $\Delta \eta$) and from the very
assumption of the pseudoplastic rheology as well, the present study
provides a fully dynamic framework in which heat flow, velocity,
lithospheric viscosity, aspect ratio, and the radius of curvature are
all connected in a self-consistent manner.

%\appendix % note: this ``appendix'' command screws up figure numbering.
\section*{Appendix A: Numerical results}
\setcounter{equation}{0}
\renewcommand{\theequation}{A\arabic{equation}}

Tabulated are selected convection diagnostics for statistically
steady-state solutions as described in the main text. The Rayleigh
number $Ra$ is $10^6$ for all cases. Reference runs refer to
calculations with the standard pseudoplastic rheology, and runs with
shallow stiffening refers to those with additional depth-dependent
viscosity.

\acknowledgements This work was sponsored by the U.S. National Science
Foundation under grant EAR-0449517 and Microsoft A. Richard Newton
Breakthrough Research Award. The author thanks the Associate Editor
and two anonymous referees for careful reviews. 

\bibliographystyle{agufull} 
\bibliography{refs}

%\setcounter{page}{31}
%
% tables
%

\begin{table*}[t!]
\scriptsize
\singlespacing
\tablenum{A1}
\begin{center}
\caption{Numerical results for reference runs ($\gamma <0.5$)}
\label{table:runKa}
\begin{tabular*}{42pc}{@{\extracolsep{\fill}}cccccccccccccc}
\tableline
$\gamma$ & $\theta_0$ & $H^*$ & $T^*_{\mbox {\scriptsize max}}$ & $T^*_i$
& $\langle T^* \rangle$ & $Nu$ & $\theta$ & $Ra_i$ 
& $v^*_{\mbox {\scriptsize rms}}$ & $v^*_s$ & $P_{0.1}$ & $\Phi$ & $\Phi_{\delta}/\Phi$ \\
\hline \\ [-2ex]
0.1 & 15 & 8 & 0.86 & 0.59 & 0.60 & 9.27 & 12.9 & $1.10 \times 10^{5}$ & $6.40 \times 10^{1}$ & $1.49 \times 10^{2}$ & 0.33 & $2.73 \times 10^{7}$ & 0.19 \\ 
0.1 & 15 & 12 & 0.93 & 0.68 & 0.69 & 12.93 & 13.9 & $3.18 \times 10^{5}$ & $1.12 \times 10^{2}$ & $2.26 \times 10^{2}$ & 0.30 & $4.25 \times 10^{7}$ & 0.14 \\ 
0.1 & 15 & 16 & 0.98 & 0.75 & 0.76 & 16.39 & 14.6 & $6.85 \times 10^{5}$ & $1.68 \times 10^{2}$ & $3.10 \times 10^{2}$ & 0.26 & $5.77 \times 10^{7}$ & 0.13 \\ 
0.1 & 15 & 20 & 1.02 & 0.82 & 0.81 & 19.67 & 15.3 & $1.32 \times 10^{6}$ & $2.33 \times 10^{2}$ & $3.97 \times 10^{2}$ & 0.22 & $7.31 \times 10^{7}$ & 0.14 \\ 
0.1 & 20 & 12 & 0.96 & 0.71 & 0.70 & 12.50 & 19.2 & $4.36 \times 10^{5}$ & $1.00 \times 10^{2}$ & $2.22 \times 10^{2}$ & 0.27 & $4.22 \times 10^{7}$ & 0.16 \\ 
0.1 & 20 & 16 & 1.00 & 0.77 & 0.77 & 15.95 & 20.1 & $1.08 \times 10^{6}$ & $1.57 \times 10^{2}$ & $3.18 \times 10^{2}$ & 0.28 & $5.77 \times 10^{7}$ & 0.13 \\
\end{tabular*}
\end{center}
\end{table*}

\begin{table*}[t!]
\scriptsize
\singlespacing
\tablenum{A2}
\begin{center}
\caption{Numerical results for reference runs ($\gamma > 0.5$)}
\label{table:runKb}
\begin{tabular*}{42pc}{@{\extracolsep{\fill}}cccccccccccccc}
\tableline
$\gamma$ & $\theta_0$ & $H^*$ & $T^*_{\mbox {\scriptsize max}}$ & $T^*_i$
& $\langle T^* \rangle$ & $Nu$ & $\theta$ & $Ra_i$ 
& $v^*_{\mbox {\scriptsize rms}}$ & $v^*_s$ & $P_{0.1}$ & $\Phi$ & $\Phi_{\delta}/\Phi$ \\
\hline \\ [-2ex]
0.6 & 10 & 8 & 0.95 & 0.64 & 0.70 & 8.39 & 9.5 & $6.05 \times 10^{5}$ & $9.53 \times 10^{1}$ & $1.36 \times 10^{2}$ & 0.51 & $2.68 \times 10^{7}$ & 0.34 \\ 
0.6 & 10 & 12 & 1.04 & 0.71 & 0.79 & 11.55 & 10.4 & $1.54 \times 10^{6}$ & $1.66 \times 10^{2}$ & $2.41 \times 10^{2}$ & 0.56 & $4.23 \times 10^{7}$ & 0.29 \\ 
0.6 & 10 & 16 & 1.10 & 0.76 & 0.86 & 14.52 & 11.0 & $3.07 \times 10^{6}$ & $2.42 \times 10^{2}$ & $3.56 \times 10^{2}$ & 0.56 & $5.78 \times 10^{7}$ & 0.28 \\ 
0.6 & 10 & 20 & 1.16 & 0.80 & 0.92 & 17.31 & 11.6 & $5.50 \times 10^{6}$ & $3.15 \times 10^{2}$ & $4.63 \times 10^{2}$ & 0.55 & $7.34 \times 10^{7}$ & 0.28 \\ 
0.6 & 15 & 8 & 1.10 & 0.74 & 0.76 & 7.29 & 16.5 & $4.76 \times 10^{6}$ & $6.65 \times 10^{1}$ & $1.03 \times 10^{2}$ & 0.36 & $2.61 \times 10^{7}$ & 0.45 \\ 
0.6 & 15 & 12 & 1.13 & 0.79 & 0.85 & 10.60 & 17.0 & $8.32 \times 10^{6}$ & $1.92 \times 10^{2}$ & $2.90 \times 10^{2}$ & 0.69 & $4.17 \times 10^{7}$ & 0.30 \\
\end{tabular*}
\end{center}
\end{table*}

\begin{table*}[t!]
\scriptsize
\singlespacing
\tablenum{A3}
\begin{center}
\caption{Numerical results with shallow stiffening ($\gamma = 0.4$, $\theta_0 = 15$,
and $H^*>7$)}
\label{table:runLa1}
\begin{tabular*}{42pc}{@{\extracolsep{\fill}}cccccccccccccc}
\tableline
$\theta_0$ & $H^*$ & $h^*$ & $\Delta \eta$ & $T^*_{\mbox {\scriptsize max}}$ 
& $\langle T^* \rangle$ & $Nu$ & $\theta$ & $Ra_i$ 
& $v^*_{\mbox {\scriptsize rms}}$ & $v^*_s$ & $P_{0.1}$ & $\Phi$ & $\Phi_{\delta}/\Phi$ \\
\hline \\ [-2ex]
15 & 8 & 0.1 & 3 & 1.08 & 0.73 & 7.43 & 16.1 & $3.38 \times 10^{6}$ & $6.49 \times 10^{1}$ & $1.10 \times 10^{2}$ & 0.59 & $2.64 \times 10^{7}$ & 0.38 \\ 
15 & 8 & 0.1 & 10 & 1.06 & 0.75 & 7.53 & 15.9 & $2.74 \times 10^{6}$ & $6.67 \times 10^{1}$ & $9.24 \times 10^{1}$ & 0.71 & $2.62 \times 10^{7}$ & 0.33 \\ 
15 & 8 & 0.1 & 30 & 1.05 & 0.75 & 7.62 & 15.8 & $2.28 \times 10^{6}$ & $7.51 \times 10^{1}$ & $1.10 \times 10^{2}$ & 0.69 & $2.62 \times 10^{7}$ & 0.33 \\ 
15 & 8 & 0.1 & 100 & 1.07 & 0.75 & 7.49 & 16.0 & $3.01 \times 10^{6}$ & $7.62 \times 10^{1}$ & $1.13 \times 10^{2}$ & 0.67 & $2.62 \times 10^{7}$ & 0.33 \\ 
15 & 8 & 0.2 & 3 & 1.05 & 0.76 & 7.63 & 15.7 & $2.22 \times 10^{6}$ & $9.88 \times 10^{1}$ & $1.41 \times 10^{2}$ & 0.65 & $2.60 \times 10^{7}$ & 0.33 \\ 
15 & 8 & 0.2 & 10 & 1.08 & 0.79 & 7.40 & 16.2 & $3.69 \times 10^{6}$ & $1.07 \times 10^{2}$ & $1.43 \times 10^{2}$ & 0.71 & $2.59 \times 10^{7}$ & 0.33 \\
\end{tabular*}
\end{center}
\end{table*}

\begin{table*}[t!]
\scriptsize
\singlespacing
\tablenum{A4}
\begin{center}
\caption{Numerical results with shallow stiffening ($\gamma = 0.4$, $\theta_0 = 20$, 
and $H^*>7$)}
\label{table:runLa2}
\begin{tabular*}{42pc}{@{\extracolsep{\fill}}cccccccccccccc}
\tableline
$\theta_0$ & $H^*$ & $h^*$ & $\Delta \eta$ & $T^*_{\mbox {\scriptsize max}}$ 
& $\langle T^* \rangle$ & $Nu$ & $\theta$ & $Ra_i$ 
& $v^*_{\mbox {\scriptsize rms}}$ & $v^*_s$ & $P_{0.1}$ & $\Phi$ & $\Phi_{\delta}/\Phi$ \\
\hline \\ [-2ex]
20 & 8 & 0.1 & 3 & 1.06 & 0.74 & 7.52 & 21.3 & $3.80 \times 10^{6}$ & $5.14 \times 10^{1}$ & $7.50 \times 10^{1}$ & 0.51 & $2.58 \times 10^{7}$ & 0.34 \\ 
20 & 8 & 0.1 & 10 & 1.08 & 0.74 & 7.42 & 21.6 & $5.10 \times 10^{6}$ & $5.77 \times 10^{1}$ & $9.57 \times 10^{1}$ & 0.50 & $2.57 \times 10^{7}$ & 0.37 \\ 
20 & 8 & 0.1 & 30 & 1.09 & 0.75 & 7.32 & 21.9 & $7.02 \times 10^{6}$ & $5.60 \times 10^{1}$ & $9.39 \times 10^{1}$ & 0.49 & $2.56 \times 10^{7}$ & 0.38 \\ 
20 & 8 & 0.1 & 100 & 1.12 & 0.76 & 7.11 & 22.5 & $1.37 \times 10^{7}$ & $5.59 \times 10^{1}$ & $8.57 \times 10^{1}$ & 0.52 & $2.52 \times 10^{7}$ & 0.38 \\ 
20 & 8 & 0.2 & 3 & 1.10 & 0.75 & 7.28 & 22.0 & $7.94 \times 10^{6}$ & $5.74 \times 10^{1}$ & $9.07 \times 10^{1}$ & 0.47 & $2.57 \times 10^{7}$ & 0.37 \\ 
20 & 8 & 0.2 & 10 & 1.13 & 0.78 & 7.08 & 22.6 & $1.52 \times 10^{7}$ & $5.34 \times 10^{1}$ & $7.60 \times 10^{1}$ & 0.50 & $2.57 \times 10^{7}$ & 0.34 \\
\end{tabular*}
\end{center}
\end{table*}

\begin{table*}[t!]
\scriptsize
\singlespacing
\tablenum{A5}
\begin{center}
\caption{Numerical results with shallow stiffening ($\gamma = 0.4$, $\theta_0 = 25$, 
and $H^*>7$)}
\label{table:runLa3}
\begin{tabular*}{42pc}{@{\extracolsep{\fill}}cccccccccccccc}
\tableline
$\theta_0$ & $H^*$ & $h^*$ & $\Delta \eta$ & $T^*_{\mbox {\scriptsize max}}$ 
& $\langle T^* \rangle$ & $Nu$ & $\theta$ & $Ra_i$ 
& $v^*_{\mbox {\scriptsize rms}}$ & $v^*_s$ & $P_{0.1}$ & $\Phi$ & $\Phi_{\delta}/\Phi$ \\
\hline \\ [-2ex]
25 & 8 & 0.1 & 3 & 1.09 & 0.76 & 7.37 & 27.1 & $9.32 \times 10^{6}$ & $6.88 \times 10^{1}$ & $1.07 \times 10^{2}$ & 0.49 & $2.46 \times 10^{7}$ & 0.36 \\ 
25 & 8 & 0.1 & 10 & 1.29 & 1.16 & 6.20 & 32.3 & $1.86 \times 10^{9}$ & $2.99 \times 10^{3}$ & $6.71 \times 10^{-4}$ & 0.00 & $2.05 \times 10^{7}$ & 0.43 \\ 
25 & 8 & 0.1 & 30 & 1.29 & 1.16 & 6.22 & 32.2 & $1.67 \times 10^{9}$ & $3.02 \times 10^{3}$ & $9.93 \times 10^{-5}$ & 0.00 & $2.06 \times 10^{7}$ & 0.43 \\ 
25 & 8 & 0.1 & 100 & 1.11 & 0.80 & 7.20 & 27.8 & $1.87 \times 10^{7}$ & $8.04 \times 10^{1}$ & $9.53 \times 10^{1}$ & 0.55 & $2.40 \times 10^{7}$ & 0.34 \\ 
25 & 8 & 0.2 & 3 & 1.11 & 0.77 & 7.22 & 27.7 & $1.69 \times 10^{7}$ & $6.59 \times 10^{1}$ & $9.62 \times 10^{1}$ & 0.49 & $2.48 \times 10^{7}$ & 0.36 \\ 
25 & 8 & 0.2 & 10 & 1.14 & 0.80 & 7.00 & 28.6 & $4.12 \times 10^{7}$ & $7.04 \times 10^{1}$ & $9.08 \times 10^{1}$ & 0.54 & $2.45 \times 10^{7}$ & 0.35 \\
\end{tabular*}
\end{center}
\end{table*}

\begin{table*}[t!]
\scriptsize
\singlespacing
\tablenum{A6}
\begin{center}
\caption{Numerical results with shallow stiffening ($\gamma = 0.4$, $\theta_0 = 15$, 
and $H^*<7$)}
\label{table:runLb1}
\begin{tabular*}{42pc}{@{\extracolsep{\fill}}cccccccccccccc}
\tableline
$\theta_0$ & $H^*$ & $h^*$ & $\Delta \eta$ & $T^*_{\mbox {\scriptsize max}}$ 
& $\langle T^* \rangle$ & $Nu$ & $\theta$ & $Ra_i$ 
& $v^*_{\mbox {\scriptsize rms}}$ & $v^*_s$ & $P_{0.1}$ & $\Phi$ & $\Phi_{\delta}/\Phi$ \\
\hline \\ [-2ex]
15 & 2 & 0.1 & 3 & 0.82 & 0.51 & 2.44 & 12.3 & $5.59 \times 10^{4}$ & $7.67 \times 10^{0}$ & $5.19 \times 10^{0}$ & 0.15 & $2.44 \times 10^{6}$ & 0.88 \\ 
15 & 2 & 0.1 & 10 & 0.82 & 0.51 & 2.44 & 12.3 & $5.59 \times 10^{4}$ & $7.67 \times 10^{0}$ & $5.18 \times 10^{0}$ & 0.20 & $2.44 \times 10^{6}$ & 0.88 \\ 
15 & 2 & 0.1 & 30 & 0.82 & 0.51 & 2.44 & 12.3 & $5.59 \times 10^{4}$ & $7.67 \times 10^{0}$ & $5.17 \times 10^{0}$ & 0.25 & $2.44 \times 10^{6}$ & 0.88 \\ 
15 & 2 & 0.1 & 100 & 0.82 & 0.51 & 2.44 & 12.3 & $5.59 \times 10^{4}$ & $7.67 \times 10^{0}$ & $5.19 \times 10^{0}$ & 0.27 & $2.44 \times 10^{6}$ & 0.88 \\ 
15 & 2 & 0.2 & 3 & 0.83 & 0.52 & 2.41 & 12.4 & $6.44 \times 10^{4}$ & $7.88 \times 10^{0}$ & $4.85 \times 10^{0}$ & 0.35 & $2.42 \times 10^{6}$ & 0.88 \\ 
15 & 2 & 0.2 & 10 & 0.84 & 0.52 & 2.38 & 12.6 & $7.50 \times 10^{4}$ & $8.17 \times 10^{0}$ & $4.80 \times 10^{0}$ & 0.35 & $2.54 \times 10^{6}$ & 0.88 \\
\end{tabular*}
\end{center}
\end{table*}

\begin{table*}[t!]
\scriptsize
\singlespacing
\tablenum{A7}
\begin{center}
\caption{Numerical results with shallow stiffening ($\gamma = 0.4$, $\theta_0 = 20$ 
\& $25$, and $H^*<7$)}
\label{table:runLb2}
\begin{tabular*}{42pc}{@{\extracolsep{\fill}}cccccccccccccc}
\tableline
$\theta_0$ & $H^*$ & $h^*$ & $\Delta \eta$ & $T^*_{\mbox {\scriptsize max}}$ 
& $\langle T^* \rangle$ & $Nu$ & $\theta$ & $Ra_i$ 
& $v^*_{\mbox {\scriptsize rms}}$ & $v^*_s$ & $P_{0.1}$ & $\Phi$ & $\Phi_{\delta}/\Phi$ \\
\hline \\ [-2ex]
20 & 2 & 0.1 & 3 & 0.96 & 0.65 & 2.08 & 19.3 & $4.62 \times 10^{5}$ & $9.88 \times 10^{0}$ & $2.53 \times 10^{-3}$ & 0.00 & $7.02 \times 10^{5}$ & 0.94 \\ 
20 & 2 & 0.2 & 3 & 0.96 & 0.65 & 2.08 & 19.3 & $4.62 \times 10^{5}$ & $9.89 \times 10^{0}$ & $1.02 \times 10^{-3}$ & 0.00 & $7.02 \times 10^{5}$ & 0.94 \\ 
20 & 2 & 0.3 & 3 & 0.96 & 0.65 & 2.08 & 19.3 & $4.66 \times 10^{5}$ & $9.90 \times 10^{0}$ & $8.63 \times 10^{-4}$ & 0.00 & $6.98 \times 10^{5}$ & 0.94 \\ 
20 & 4 & 0.1 & 3 & 1.01 & 0.66 & 3.97 & 20.2 & $1.18 \times 10^{6}$ & $2.28 \times 10^{1}$ & $2.09 \times 10^{1}$ & 0.56 & $9.79 \times 10^{6}$ & 0.58 \\ 
20 & 4 & 0.1 & 10 & 0.98 & 0.64 & 4.07 & 19.7 & $7.25 \times 10^{5}$ & $2.21 \times 10^{1}$ & $4.45 \times 10^{1}$ & 0.37 & $9.86 \times 10^{6}$ & 0.53 \\ 
20 & 4 & 0.1 & 30 & 1.01 & 0.67 & 3.94 & 20.3 & $1.35 \times 10^{6}$ & $2.33 \times 10^{1}$ & $2.03 \times 10^{1}$ & 0.58 & $9.66 \times 10^{6}$ & 0.57 \\
\end{tabular*}
\end{center}
\end{table*}

\begin{table*}[t!]
\scriptsize
\singlespacing
\tablenum{A8}
\begin{center}
\caption{Numerical results with shallow stiffening ($\gamma = 0.8$)}
\label{table:runLe}
\begin{tabular*}{42pc}{@{\extracolsep{\fill}}cccccccccccccc}
\tableline
$\theta_0$ & $H^*$ & $h^*$ & $\Delta \eta$ & $T^*_{\mbox {\scriptsize max}}$ 
& $\langle T^* \rangle$ & $Nu$ & $\theta$ & $Ra_i$ 
& $v^*_{\mbox {\scriptsize rms}}$ & $v^*_s$ & $P_{0.1}$ & $\Phi$ & $\Phi_{\delta}/\Phi$ \\
\hline \\ [-2ex]
10 & 2 & 0.2 & 1000 & 0.85 & 0.58 & 2.35 & 8.5 & $1.91 \times 10^{5}$ & $1.79 \times 10^{1}$ & $3.66 \times 10^{-2}$ & 0.00 & $2.07 \times 10^{6}$ & 0.86 \\ 
10 & 2 & 0.3 & 10 & 0.80 & 0.53 & 2.50 & 8.0 & $1.07 \times 10^{5}$ & $1.37 \times 10^{1}$ & $3.83 \times 10^{0}$ & 0.46 & $2.72 \times 10^{6}$ & 0.83 \\ 
10 & 2 & 0.3 & 100 & 0.88 & 0.60 & 2.27 & 8.8 & $2.66 \times 10^{5}$ & $1.89 \times 10^{1}$ & $1.23 \times 10^{-1}$ & 0.00 & $1.75 \times 10^{6}$ & 0.87 \\ 
10 & 3 & 0.2 & 10 & 0.75 & 0.52 & 3.98 & 7.5 & $6.49 \times 10^{4}$ & $2.01 \times 10^{1}$ & $1.77 \times 10^{1}$ & 0.66 & $7.61 \times 10^{6}$ & 0.63 \\ 
10 & 3 & 0.2 & 100 & 1.00 & 0.78 & 2.99 & 10.0 & $1.03 \times 10^{6}$ & $5.80 \times 10^{1}$ & $4.43 \times 10^{-2}$ & 0.00 & $4.41 \times 10^{6}$ & 0.68 \\ 
10 & 3 & 0.3 & 10 & 0.83 & 0.58 & 3.62 & 8.3 & $1.49 \times 10^{5}$ & $2.14 \times 10^{1}$ & $1.39 \times 10^{1}$ & 0.57 & $6.95 \times 10^{6}$ & 0.70 \\
\end{tabular*}
\end{center}
\end{table*}

%
% Figure Captions
%
% note: I have to use Illustrator 10 (CS2 doesn't produce EPS files compatible
% with LaTeX!!)
\newpage

\end{article}
\end{document}